\documentclass[fleqn,usenatbib]{mnras}
\usepackage{lineno}
\usepackage{newtxtext,newtxmath}

\usepackage[T1]{fontenc}

\DeclareRobustCommand{\VAN}[3]{#2}
\let\VANthebibliography\thebibliography
\def\thebibliography{\DeclareRobustCommand{\VAN}[3]{##3}\VANthebibliography}

\usepackage{graphicx}	
\usepackage{amsmath}	

\usepackage{eso-pic}

\AddToShipoutPictureBG*{%
  \AtPageUpperLeft{%
    \hspace{0.75\paperwidth}%
    \raisebox{-4.5\baselineskip}{%
      \makebox[0pt][l]{\textnormal{DES-2021-0642}}
}}}%

\AddToShipoutPictureBG*{%
  \AtPageUpperLeft{%
    \hspace{0.75\paperwidth}%
    \raisebox{-5.5\baselineskip}{%
      \makebox[0pt][l]{\textnormal{FERMILAB-PUB-21-245-AE}}
}}}%

\defcitealias{Childress2014}{C14}

\title[Rates of SNe Ia in DES]{Rates and delay times of type Ia supernovae in the Dark Energy Survey}

\author[P. Wiseman et al.]{\parbox{\textwidth}{
\Large
P.~Wiseman,$^{1}$
M.~Sullivan,$^{1}$
M.~Smith,$^{2,1}$
C.~Frohmaier,$^{3,1}$
M.~Vincenzi,$^{3}$
O.~Graur,$^{3}$
B.~Popovic,$^{4}$
P.~Armstrong,$^{5}$
D.~Brout,$^{6,7}$
T.~M.~Davis,$^{8}$
L.~Galbany,$^{9}$
S.~R.~Hinton,$^{8}$
L.~Kelsey,$^{1}$
R.~Kessler,$^{10,11}$
C.~Lidman,$^{12,5}$
A.~M\"oller,$^{13}$
R.~C.~Nichol,$^{3}$
B.~Rose,$^{4}$
D.~Scolnic,$^{4}$
M.~Toy,$^{1}$
Z.~Zontou,$^{1}$
J.~Asorey,$^{14}$
D.~Carollo,$^{15}$
K.~Glazebrook,$^{16}$
G.~F.~Lewis,$^{17}$
B.~E.~Tucker,$^{5}$
T.~M.~C.~Abbott,$^{18}$
M.~Aguena,$^{19}$
S.~Allam,$^{20}$
F.~Andrade-Oliveira,$^{21,19}$
J.~Annis,$^{20}$
D.~Bacon,$^{3}$
E.~Bertin,$^{22,23}$
D.~Brooks,$^{24}$
E.~Buckley-Geer,$^{10,20}$
D.~L.~Burke,$^{25,26}$
A.~Carnero~Rosell,$^{27,19,28}$
M.~Carrasco~Kind,$^{29,30}$
J.~Carretero,$^{31}$
M.~Costanzi,$^{32,33,34}$
L.~N.~da Costa,$^{19,35}$
M.~E.~S.~Pereira,$^{36}$
S.~Desai,$^{37}$
H.~T.~Diehl,$^{20}$
P.~Doel,$^{24}$
S.~Everett,$^{38}$
I.~Ferrero,$^{39}$
B.~Flaugher,$^{20}$
P.~Fosalba,$^{40,41}$
J.~Frieman,$^{20,11}$
J.~Garc\'ia-Bellido,$^{42}$
E.~Gaztanaga,$^{40,41}$
T.~Giannantonio,$^{43,44}$
D.~Gruen,$^{45,25,26}$
R.~A.~Gruendl,$^{29,30}$
J.~Gschwend,$^{19,35}$
G.~Gutierrez,$^{20}$
D.~L.~Hollowood,$^{38}$
K.~Honscheid,$^{46,47}$
B.~Hoyle,$^{48,49}$
D.~J.~James,$^{7}$
E.~Krause,$^{50}$
K.~Kuehn,$^{51,52}$
N.~Kuropatkin,$^{20}$
M.~A.~G.~Maia,$^{19,35}$
J.~L.~Marshall,$^{53}$
P.~Martini,$^{46,54,55}$
F.~Menanteau,$^{29,30}$
R.~Miquel,$^{56,31}$
R.~Morgan,$^{57}$
R.~L.~C.~Ogando,$^{19,35}$
A.~Palmese,$^{20,11}$
F.~Paz-Chinch\'{o}n,$^{29,43}$
D.~Petravick,$^{29}$
A.~Pieres,$^{19,35}$
A.~A.~Plazas~Malag\'on,$^{58}$
A.~K.~Romer,$^{59}$
E.~Sanchez,$^{14}$
V.~Scarpine,$^{20}$
M.~Schubnell,$^{36}$
S.~Serrano,$^{40,41}$
I.~Sevilla-Noarbe,$^{14}$
M.~Soares-Santos,$^{36}$
E.~Suchyta,$^{60}$
M.~E.~C.~Swanson,$^{29}$
G.~Tarle,$^{36}$
D.~Thomas,$^{3}$
C.~To,$^{45,25,26}$
T.~N.~Varga,$^{49,61}$
and A.~R.~Walker$^{18}$
\begin{center} (DES Collaboration) \end{center}
}
\vspace{0.4cm}
\\
\parbox{\textwidth}{Affiliations are listed at the end of the paper}
}
\date{Accepted XXX. Received YYY; in original form ZZZ}

\pubyear{2020}

\begin{document}
\label{firstpage}
\pagerange{\pageref{firstpage}--\pageref{lastpage}}
\maketitle

\begin{abstract}
We use a sample of 809 photometrically classified type Ia supernovae (SNe Ia) discovered by the Dark Energy Survey (DES) along with 40415 field galaxies to calculate the rate of SNe Ia per galaxy in the redshift range $0.2 < z <0.6$. We recover the known correlation between SN Ia rate and galaxy stellar mass across a broad range of scales $8.5 \leq \log(M_*/\mathrm{M}_{\odot}) \leq 11.25$. We find that the SN Ia rate increases with stellar mass as a power-law with index $0.63 \pm 0.02$, which is consistent with previous work. We use an empirical model of stellar mass assembly to estimate the average star-formation histories (SFHs) of galaxies across the stellar mass range of our measurement. Combining the modelled SFHs with the SN Ia rates to estimate constraints on the SN Ia delay time distribution (DTD), we find the data are fit well by a power-law DTD with slope index $\beta = -1.13 \pm 0.05$ and normalisation $A = 2.11 \pm0.05 \times 10^{-13}~\mathrm{SNe}~{\mathrm{M}_{\odot}}^{-1}~\mathrm{yr}^{-1}$, which corresponds to an overall SN Ia production efficiency $N_{\mathrm{Ia}}/M_* = 0.9~_{-0.7}^{+4.0} \times 10^{-3}~\mathrm{SNe}~\mathrm{M}_{\odot}^{-1}$. Upon splitting the SN sample by properties of the light curves, we find a strong dependence on DTD slope with the SN decline rate, with slower-declining SNe exhibiting a steeper DTD slope. We interpret this as a result of a relationship between intrinsic luminosity and progenitor age, and explore the implications of the result in the context of SN Ia progenitors.
\end{abstract}




\section{Introduction}

Type Ia supernovae (SNe Ia) are explosions of white dwarf stars (WDs). Although SNe Ia show diversity in their observed properties, a large fraction of them ("non-peculiar" SNe Ia) display a small dispersion in their peak brightnesses which can be reduced further through empirical relations between brightness and light curve properties, such as decline rate (stretch) or optical colour \citep{Rust1974,Pskovskii1977,Phillips1993,Tripp1998}. These properties have led SNe Ia to be used extensively by cosmologists to measure relative distances in the Universe \citep{Riess1998,Perlmutter1999}.

Despite using these relations to correct SN Ia peak brightnesses to within a dispersion of $\sim 0.1$ mag in samples with over one thousand SNe \citep{Scolnic2018}, the exact nature of the progenitors of SNe Ia is yet to be confirmed. While it is likely that SNe Ia are caused by mass transfer onto a WD from, or violent merger with, a companion star, multiple possible scenarios exist for the nature of that companion star (see \citealt{Maoz2014,Ruiter2020} for reviews). The leading models involve a main sequence (MS) star in the single-degenerate (SD) scenario \citep{Whelan1973,Nomoto1982}, a secondary white dwarf in the double-degenerate (DD) scenario \citep{Tutukov1976,Iben1984,Webbink1984}, or more exotic models such as the core-degenerate (CD) scenario which invoke WDs merging with the cores of massive stars \citep{Kashi2011,Ilkov2011}. 

The progenitors of several core-collapse SNe (CCSNe) have been identified in high-resolution pre-explosion images \citep{Smartt2009,Eldridge2013}, allowing detailed analysis of the latter stages of massive star evolution. However, this approach has not yet been successful in identifying a progenitor of a SN Ia (e.g. \citealt{Graur2014a,Kelly2014,Graur2019}, although see \citet{McCully2014} for evidence of a progenitor for a type Iax SN). Conversely, there have been searches for surviving remnants of the binary system -- main sequence stars or WDs left behind at the SN location or kicked out at high velocity \citep[e.g.][]{Schaefer2012,Ruiz-Lapuente2018,Kerzendorf2018,Kerzendorf2019}. Promising recent results include that from \citet{Shen2018} who discovered three high-velocity WDs consistent with models of double-degenerate, double-detonation SNe Ia. However, without an unambiguous observation of a SN Ia progenitor system or remnant companion, there is as yet no direct evidence that any progenitor channel from provides a significant contribution to the total population of SNe Ia.

An indirect method of inferring progenitor channels is the study of the SN Ia delay time distribution (DTD). The DTD describes the rate at which SNe Ia occur as a function of the delay time $\tau$ since an episode of star-formation, and thus carries characteristic signatures of the progenitor channel (see \citealt{Wang2012} for a review of the theory, and \citealt{Maoz2017} for a review of observations). The SD scenario produces a broad range of functional forms for the DTD, most of which fail to account for long delay times \citep{Graur2014}. On the other hand, most variants of the DD scenario predict a power-law of the form $\tau^{\beta}$, with $\beta \sim -1$ \citep[e.g][]{Ruiter2009,Mennekens2010}, although see e.g.  \citet{Yungelson2017} for DD models that deviate from a $-1$ power-law slope.

As a statistical distribution acting on timescales from tens of Myr to several Gyr after an epoch of star formation, the DTD is non-trivial to measure and various techniques have been developed to infer it indirectly. One such approach is to measure global properties of SN Ia host galaxies. Simple examples of such analyses are comparisons of the rates of SNe Ia in hosts that have been split by some observational property, such as morphology or colour, and have shown that SN Ia rate per unit stellar mass (SNuM) is significantly larger in late-type spiral galaxies as well as in bluer galaxies \citep[e.g.][]{Mannucci2005}. These observations were interpreted as showing that SNe Ia are strongly influenced by recent or ongoing star-formation, and thus that the majority of SNe Ia explode after short delay times. On the other hand, the SNuM in E/S0 and red galaxies is also non-negligible, suggesting that there is a secondary, much older component to the DTD \citep{Sullivan2006,Li2011a,Smith2012}. The DTD was thus approximated to first order by a two-component ("$A + B$") model, where $A$ and $B$ are the normalisations of the DTD for "prompt" (i.e. proportional to instantaneous SFR) and "tardy" (i.e. proportional to overall stellar mass) SNe, respectively.

A complementary technique involves measuring the evolution of the volumetric rate of SNe Ia as a function of redshift and comparing this evolution to the average cosmic star-formation history (CSFH) of the Universe \citep{Gal-Yam2004,Strolger2004,Dahlen2004,Dahlen2008, Graur2011,Graur2013, Rodney2014,Frohmaier2019}. This technique (known as the volumetric rate method) is also applicable to galaxy clusters, for which it is assumed that the star-formation histories are strongly peaked at some past epoch and thus that the rate of SNe Ia in clusters as a function of redshift is a more direct measure of the DTD \citep{Maoz2004,Maoz2010,Friedmann2018,Freundlich2021}. These studies have, almost ubiquitously, found $\beta$ to be consistent with $-1$.

Instead of comparing \textit{volumetric} rates to the \textit{cosmic} SFH, it is also possible to estimate the SFH of individual galaxies through the modelling of their stellar populations via spectral energy distribution (SED) fitting. Works such as \citet{Totani2008}, \citet{Maoz2011}, \citet{Maoz2012}, \citet{Graur2013}, and \citet{Graur2015} estimated the DTD by measuring SFHs for a sample of field galaxies and comparing them to the number of SNe detected in each galaxy (the SFHR method), and have led to results that suggest a DTD power law with $\beta$ consistent with $-1$. 

\citet[][hereafter C14]{Childress2014} showed that the $A+B$ approximation arises as a direct consequence of the combination of a power-law DTD with the average SFHs of galaxies -- the prompt component proportional to the amount of on-going star-formation at the epoch of observation, and the tardy component caused by the fact that massive galaxies experienced high SFRs several Gyrs ago. Recent advances in integral field spectroscopy (IFS) allow for an extension of the SFHR method, by reconstructing the SFH for hundreds or thousands of local regions of each SN Ia host galaxy. Using this method, \citet{Castrillo2020} find a power-law slope of $-1.1\pm0.3$, while \citet{Chen2021} find $-1.4\pm0.3$.

The majority of observational evidence based on studies of SN Ia populations thus points towards a DD scenario for most, if not all, SNe Ia. However, finding self-consistent progenitor and explosion models that recreate the observed luminosity function as well as correlations between luminosity, light curve parameters, and host galaxy properties has proven difficult. In particular, simulations based around explosions of $M_{\mathrm{Ch}}$ WDs (linked strongly with the SD scenario but also with many DD scenarios) find difficulty in reproducing the light curves of "normal" SNe Ia as well as "peculiar" objects (\citealt{Ropke2007,Sim2013,Blondin2017}; see \citealt{Maoz2014,Jha2019} for overviews). In recent years, attention directed towards explosions of sub-$M_{\mathrm{Ch}}$ WDs triggered by double detonations (primarily related to a DD scenario) has led to promising results \citep[e.g.][]{Shen2017,Shen2018,Townsley2019,Gronow2020,Shen2021} although they still struggle to match observations at late times in the light curve evolution \citep{Shen2021,Gronow2021}. An additional factor in support of the sub-$M_{\mathrm{Ch}}$ model is that the SN luminosity is related to the mass of the primary WD, which itself is likely to be related to its age (although this relation is probably complicated by other factors such as accretion rate, metallicity, and the composition of the companion), thereby providing an explanation for observed relation between light curve stretch and stellar age \citep{Rigault2013,Rigault2018,Rose2019,Nicolas2020}. Other proposed scenarios include hybrid models in which standard CO WDs merge with hybrid helium-CO WDs \citep{Zenati2019}. With many models showing promising similarities to observations but each subject to its own drawbacks, it is becoming accepted that more than one progenitor scenario may contribute significantly to the overall population of "normal" SNe Ia; detailed observations are thus required in order to place constraints on the relative fractions of each possible progenitor channel. 

In this work, we combine parts of the traditional methods and derive a new measurement of the SN Ia DTD. Instead of measuring the volumetric rate of SNe Ia, we measure the rate per galaxy as a function of stellar mass, as per \citet{Sullivan2006,Smith2012,Brown2019}. We use the stellar mass assembly model of \citetalias{Childress2014} to predict the stellar age distribution of galaxies for a given stellar mass at the mean redshift of our SN sample, and then forward model the DTD by convolving it with the stellar age distribution and comparing the predicted rates to the observed rates.

In Section \ref{sec:data} we introduce our large sample of SNe Ia from the Dark Energy Survey (DES) as well as our deep sample of field galaxies that provide an effectively complete control sample from which to measure the rate. We describe our detailed handling of SN and galaxy incompleteness in Section \ref{sec:incompleteness}. We present the SN Ia rate per galaxy in Section \ref{sec:rates} and show that it is consistent with previous works. We outline our modelling of star-formation histories and our novel constraints on the DTD parameters in Section \ref{sec:model}. In Section \ref{sec:split_x1_c} we investigate how the DTD differs among sub-populations of SNe with different light curve characteristics, in particular the stretch, and show stretch to be strongly dependent on progenitor age. We conclude in Section \ref{sec:conclusion} by discussing the implications of the results.
Where relevant, we assume a spatially flat $\Lambda$CDM cosmology with $\Omega_m = 0.3$ and $H_0 = 70 \mathrm{~km~s}^{-1}~\mathrm{Mpc}^{-1}$. Unless otherwise stated, we assume Gaussian measurement uncertainties quoted at the $1\sigma$ level, and we quote the posterior median and $1\sigma$ credible intervals on derived parameters. Magnitudes are quoted in the AB system \citep{Oke1983}.

\begin{table}
	\centering
	\caption{Numbers of field galaxies and SN hosts passing various quality cuts. SNe are derived from must already have passed the masked chips and good redshift cuts before reaching this stage.}
	\label{tab:cuts}
	\begin{tabular}{lccr} 
		\hline
		Cut & N (field galaxies)  & N (SN hosts)\\
		\hline
		Chosen fields & 1364311  & 1441\\
	    Kron mag in all bands & 1069004  & 1439 \\
	    Masked chips & 816950  & - \\
	    Has ugrizJHK photometry & 545748  & -\\
	    Edge of chip & 481731 & 1401 \\
	    Star/galaxy separation & 400051 &  1259\\
	    Has good redshift & 395034 & - \\
	    SNR $\geq 3$& 338256  & 1259 \\
	    $m_r \leq 24.5$ & 196109 &  1254 \\
	    $0.2 \leq z \leq 0.6$ & 48177 & 809\\ 
	    
		\hline
	\end{tabular}
\end{table}

\section{Data}

In order to measure the rate of SNe Ia per galaxy per year, we require a sample of SNe Ia, as well as a sample of all of the possible galaxies ("field" galaxies) that those SNe could have exploded in. In practice, SN and galaxy surveys do not cover the same sky area, redshift ranges, times, and have different selection biases. In this section, we introduce our SN and field galaxy samples and in Section \ref{sec:incompleteness} we describe our method of correcting the selection effects for both the SN and field galaxy samples.

\label{sec:data}
\subsection{Dark Energy Survey supernova programme \label{subsec:des}}
To derive our sample of SNe and field galaxies, we make use of the Dark Energy Survey (DES). The DES Supernova Programme (DES-SN) was a transient survey based on five six-month seasons of observations of ten southern hemisphere fields with the Dark Energy Camera (DECam; \citealt{Flaugher2015}). The SN survey was designed primarily to measure the light curves of SNe Ia for use as cosmological distance indicators. Transients were detected and processed using a difference imaging pipeline \citep{Kessler2015} and image-subtraction artefacts were rejected using a machine learning (ML) algorithm \citep{Goldstein2015}. Spectral follow-up of live SN candidates was performed on a suite of large optical telescopes \citep{Smith2020a}, leading to the initial publication of a measurement of cosmological parameters using 207 spectroscopically confirmed SNe Ia (\citealt{DESCollaboration2018a}, and references therein). Spectroscopic redshifts for host galaxies come from the OzDES Global Redshift Catalog (GRC; \footnote{\url{https://docs.datacentral.org.au/ozdes/overview/dr2/}} \citealt{Yuan2015,Childress2017,Lidman2020}), which comprises galaxies targetted as DES-SN hosts with the OzDES programme as well as redshifts from legacy catalogues in the DES-SN fields.
\subsection{Supernovae \label{subsec:host_sample}}
\subsubsection{Photometric classification and quality cuts \label{subsubsec:sn_classify}}
With over 30,000 discovered transients it was not possible to obtain spectral follow up of every object. Instead we make use of photometric classification in the form of the recurrent neural network classifier \texttt{superNNova} (SNN; \citealt{Moller2019}), following the approach of \citet{Scolnic2020}. A full description of the training and fitting of SNN will be presented in Vincenzi et al. \textit{in prep} and M\"{o}ller et al. \textit{in prep}; we provide a brief overview here.

We train SNN using a large suite of simulated multi-band SN light curves and their associated host galaxy redshifts, using its default architecture. SNe Ia are simulated based on the \texttt{SALT2} model \citep{Guy2007} trained on the Joint Lightcurve Analysis (JLA) data set \citep{Betoule2014} using an identical method to that described in detail in \citet{Vincenzi2020}, and based on the workflow outlined in \citet{Kessler2019}. We simulate SNe using the SuperNova ANAlysis software (\texttt{SNANA}; \citealt{Kessler2009a}) integrated into the \texttt{pippin} framework \citep{Hinton2020}. SALT2 parameters $x_1$ (stretch) and $c$ (colour) are drawn randomly from the intrinsic asymmetric Gaussian distributions described in \citet{Scolnic2016} and "blurred" following the intrinsic scatter model of \citet{Guy2010}, while redshifts are drawn following the volumetric rate evolution of \citet{Frohmaier2019}. Synthetic CCSNe light curves are generated from the templates of \citet{Vincenzi2019}, with a rate following the CSFH of \citet{Madau2014} and normalised by the local Universe rate of \citet{Frohmaier2020}. 

Given the training set of synthetic light curves, SNN returns a classification (SN Ia, CCSN, or peculiar) for observed light curves. Before classifying, we remove transients with variability in multiple seasons (likely active galactic nuclei or superluminous SNe). SNN is then run on every transient with a host galaxy redshift (see Section \ref{subsubsec:sn_hosts}). We define SNe with a threshold probability of $P(\mathrm{Ia})\geq0.5$ as photometrically classified SNe Ia, although our analysis is not sensitive to this choice, since the vast majority of SNe receive classifications close to 1 or 0. The photometrically classified SNe Ia are then passed through a \texttt{SALT2} light curve fitting code and are subject to quality cuts on $x_1$ and $c$ in an identical manner to \citet{Vincenzi2020}: $-3 \leq x_1 \leq 3$, $-0.3 \leq c \leq 0.3$. This selection helps to reduce the potential contamination from core-collapse SNe (CCSNe) and outlying thermonuclear SNe, as well as removing those in extremely dusty environments. We impose further cuts on the quality of those measured parameters, which are: the uncertainty on $x_1$: $\sigma_{x_1} <1$; the uncertainty on the date at which the light curve has peak brightness: $\sigma_{t_{\mathrm{peak}}} <2$~d; the SALT2 fit probability\footnote{based on light curve fit $\chi^2$ and number of degrees of freedom.} $>0.01$. The number of SNe passing each stage of these cuts is displayed in Table 2 of \citet{Vincenzi2020} and reduces the $\sim30,000$ objects to 1604 SNe. With the addition of the photometric classifier we are left with 1441 SNe with which we proceed to inspect their host galaxies. While the inclusion of stringent cuts significantly reduces the size of the sample, it greatly enhances the purity and is in line with previous rates analyses of this nature \citep{Sullivan2006} as well as being consistent with other DES-SN analyses \citep[e.g.][]{Kelsey2021}. The contamination of this sample from CCSNe according to the simulations of \citet{Vincenzi2020} is expected to be below 3\%, which is close to that of spectroscopic samples \citep{Rubin2015}.

\subsubsection{Host galaxy selection \label{subsubsec:sn_hosts}}
Host galaxies for all transients in DES-SN are retrieved from the DES-SN Deep catalogue, a galaxy catalogue from the $g, r, i, z$-band coadded images of the ten DES-SN fields which was presented in \citet{Wiseman2020}. SNe are associated to galaxies using the directional light radius (DLR) method \citep[e.g.][]{Sullivan2006,Gupta2016}. As is standard across DES-SN analyses, we require a SN-host separation to DLR ratio ($d_{\mathrm{DLR}}$) less than 4 in order to classify a galaxy as a host. In the case where multiple galaxies lie within $d_{\mathrm{DLR}}<4$, the object with the lower or lowest value is taken to be the host. 

\subsection{Field galaxies\label{subsec:field_sample}}

To calculate the SN Ia rate per galaxy as a function of stellar mass we require a representative sample of the global galaxy population, which we call our field galaxy sample. In order to maintain consistency between the SN host and field galaxy sample selections, we obtain a field galaxy sample from the same catalogue as that from which the SNe are matched to obtain host galaxies (SN Deep; \citealt{Wiseman2020}). The steps relevant to this analysis are outlined in the following sections.

\subsubsection{Photometric redshifts \label{subsubsec:photozs}}

Since the vast majority of objects detected in the SN Deep catalogue do not have a spectroscopic redshift measurement, we rely on photometric redshifts (photo-$z$s).
Photo-$z$s are taken, from the DES Y3A2\_DEEP catalogue of \citet{Hartley2020}. Y3A2\_DEEP makes use of the same deep DES optical photometry that was used in SN Deep for a subset of the fields (SN-X3, SN-C3 and SN-E2), but adds DECam $u$-band data and near-infrared (NIR) $J$, $H$, and $K$-band imaging from the ultraVISTA and VIDEO surveys. The deep optical photometry in Y3A2\_DEEP was stacked using a different technique to that of SN Deep, but the resulting photometry is consistent within $1\sigma$ uncertainties (\citealt{Wiseman2020}, Meledorf et al. \textit{in prep}). \citet{Hartley2020} estimated photo-$z$s using the \texttt{eaZy} code \citep{Brammer2008}. Photo-$z$ accuracy from \citet{Hartley2020} at $17 < i<24$ is estimated around 0.03 as quantified by the Normalised Median Absolute Deviation (NMAD). This accuracy is degraded to 0.07 at $17 < i<26$, a range that includes all galaxies in our sample, but is more than adequate for the purposes of this work, where all field galaxies will be grouped into stellar mass bins of $\log \left(\Delta M_*/\mathrm{M}_{\odot}\right) = 0.25$. A detailed description of the photo-$z$ accuracy and its dependence on redshift and apparent magnitude is presented in \citet{Hartley2020}.

\begin{figure}
    \centering
    \includegraphics[width=.5\textwidth]{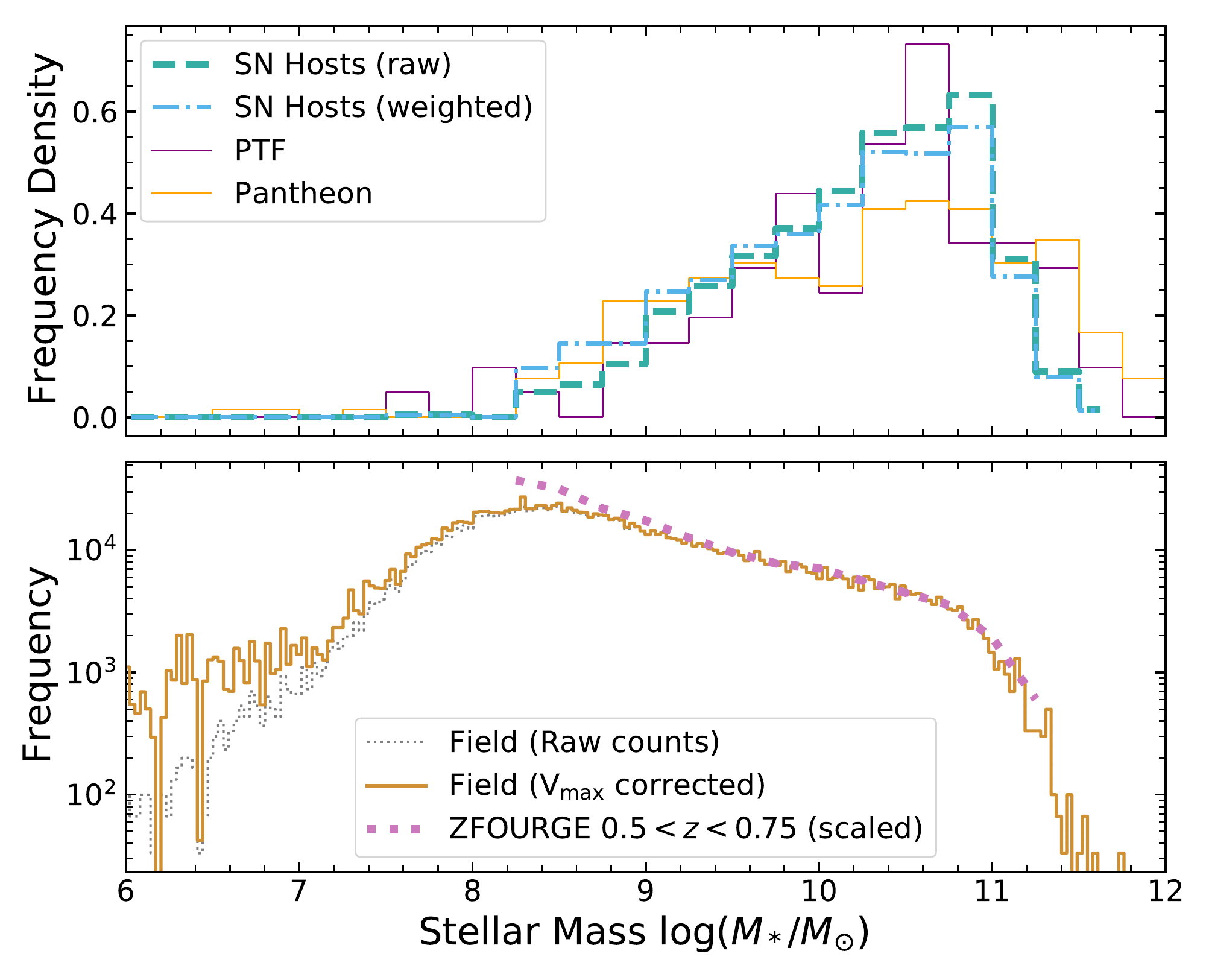}
    \caption{Upper: Distribution of the stellar mass of the SN host galaxies. The raw (green dashed) and spectroscopic-efficiency-weighted (blue dot-dashed; Section \ref{subsec:incompletenss_SN_hosts}) histograms are both normalised. Comparison samples are PTF (purple; \citealt{Pan2014}) and Pantheon (orange; \citealt{Scolnic2018}). Lower: As upper but for field galaxies, and with galaxy frequency shown on a log scale.  The distribution closely matches data from the ZFOURGE survey at $0.5<z<0.75$ (magenta dots). The difference between the raw counts (grey dotted) and the $V_{\mathrm{max}}$ corrected counts (orange solid; see Section \ref{sec:incompleteness}) is minimal. 
    \label{fig:mass_hist_hosts_field}}
\end{figure}

\begin{figure}
    \centering
    \includegraphics[width=0.5\textwidth]{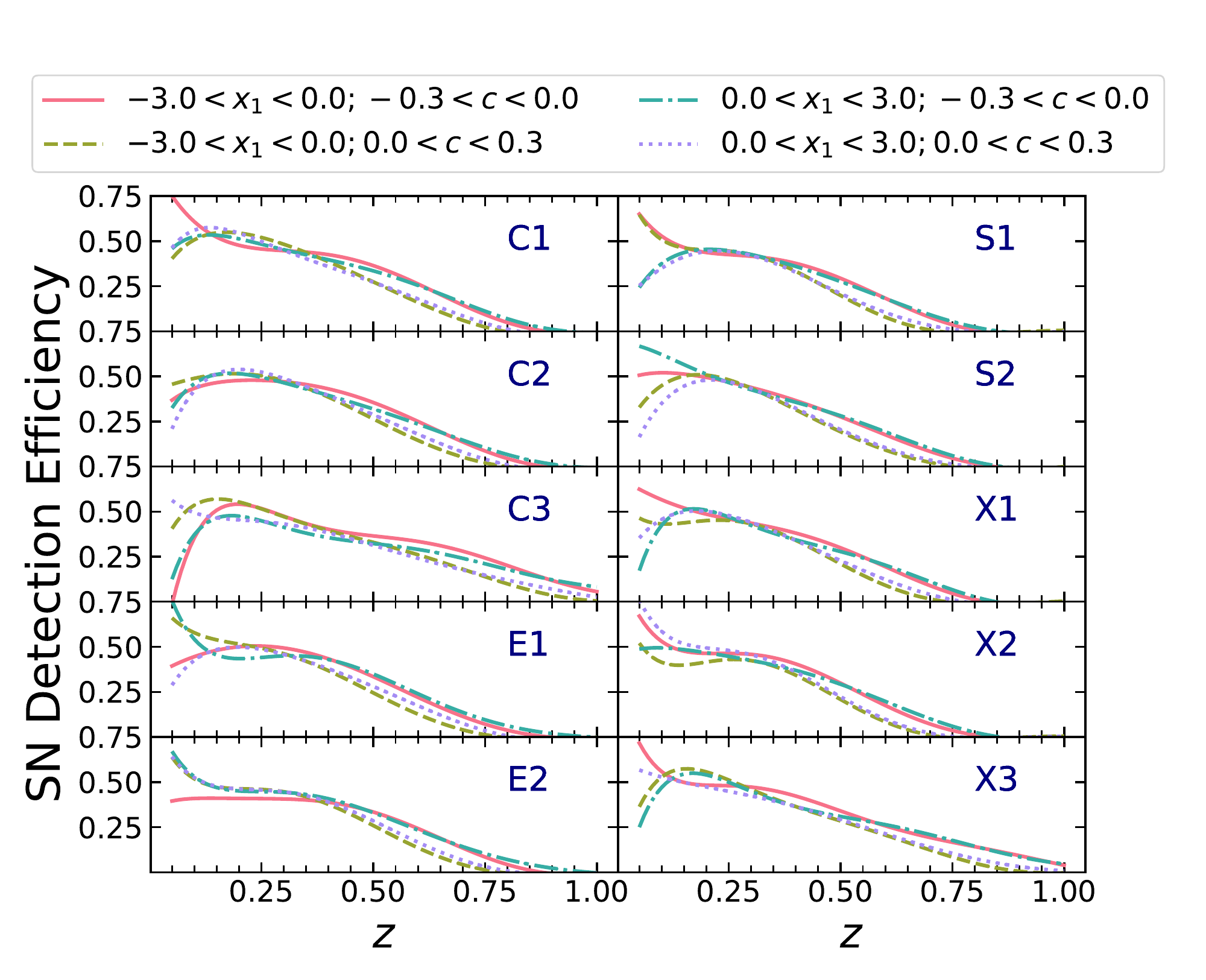}
    \caption{ SN detection efficiencies split by DES-SN field, $x_1$ and $c$. The $y-$axis represents the fraction of simulated SNe in a given redshift bin that would have been detected by DES-SN and passed light curve quality cuts. Lines are polynomial fits that approximate the efficiency curves. 
    \label{fig:SN_efficiency}}
\end{figure}

\begin{figure}
    \centering
    \includegraphics[width=0.5\textwidth]{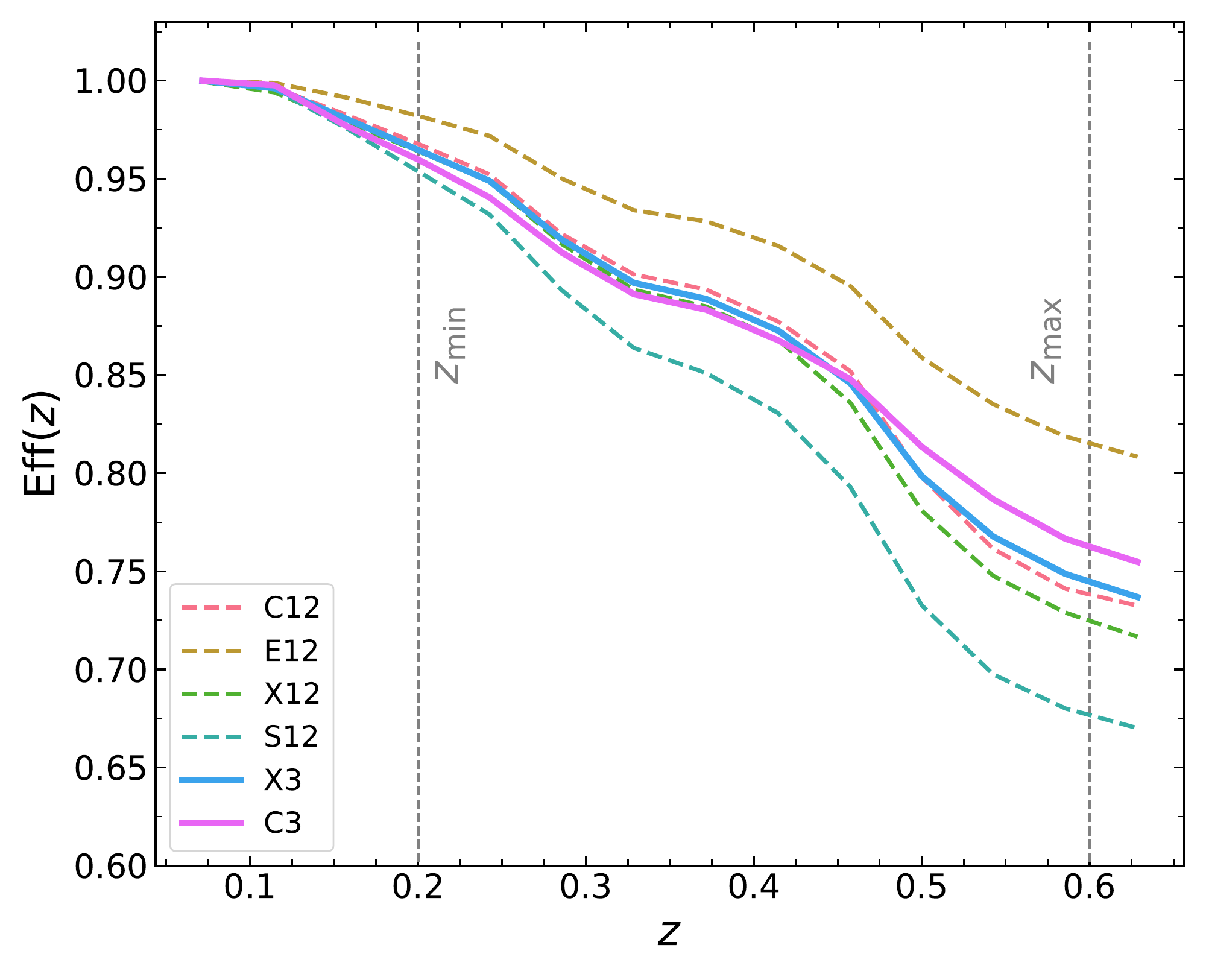}
    \caption{SN host galaxy spectroscopic efficiencies as a function of redshift $z$. Efficiency curves are grouped by shallow (\{F\}12) and deep (\{F\}3) fields, and averaged over the 5 years of the survey. The lower and upper redshift limits for this analysis are indicated by $z_{\mathrm{min}}$ and $z_{\mathrm{max}}$ respectively.
    \label{fig:eff_z}}
\end{figure}

\begin{figure}
    \centering
    \includegraphics[width=0.5\textwidth]{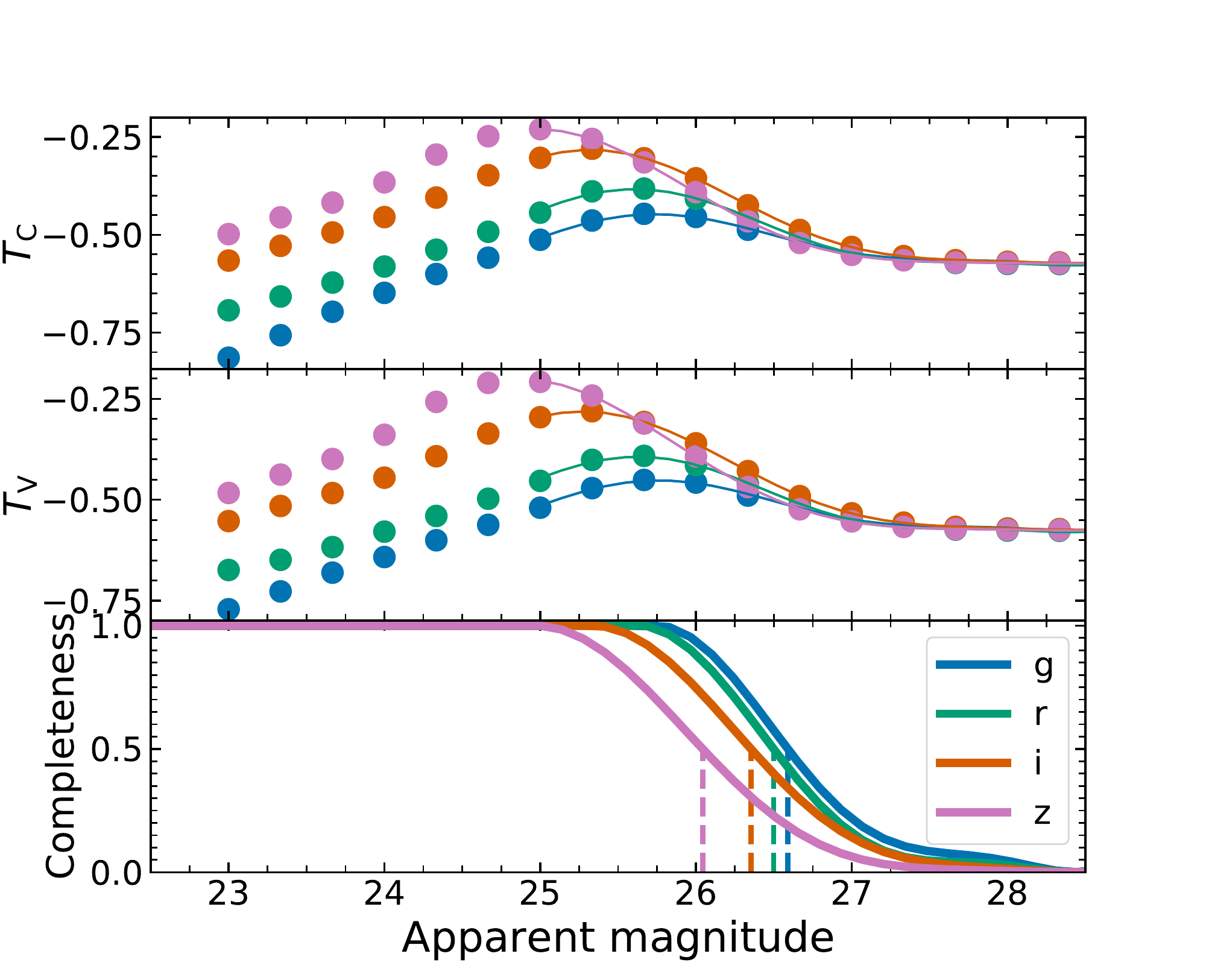}
    \caption{Measuring the completeness of the field galaxy sample. \textit{Upper}: the $T_{\mathrm{C}}$ galaxy completeness statistic (which is calculated by ranking objects by absolute magnitude in slices of distance modulus) as a function of apparent magnitude; \textit{middle} as per \textit{upper}, but for the $T_{\mathrm{V}}$ statistic which is measured by ranking galaxies by distance modulus in slices of absolute magnitude; \textit{lower}: the combined, normalised completeness in each band.  
    \label{fig:completeness}}
\end{figure}

\begin{figure}
    \centering
    \includegraphics[width=0.5\textwidth]{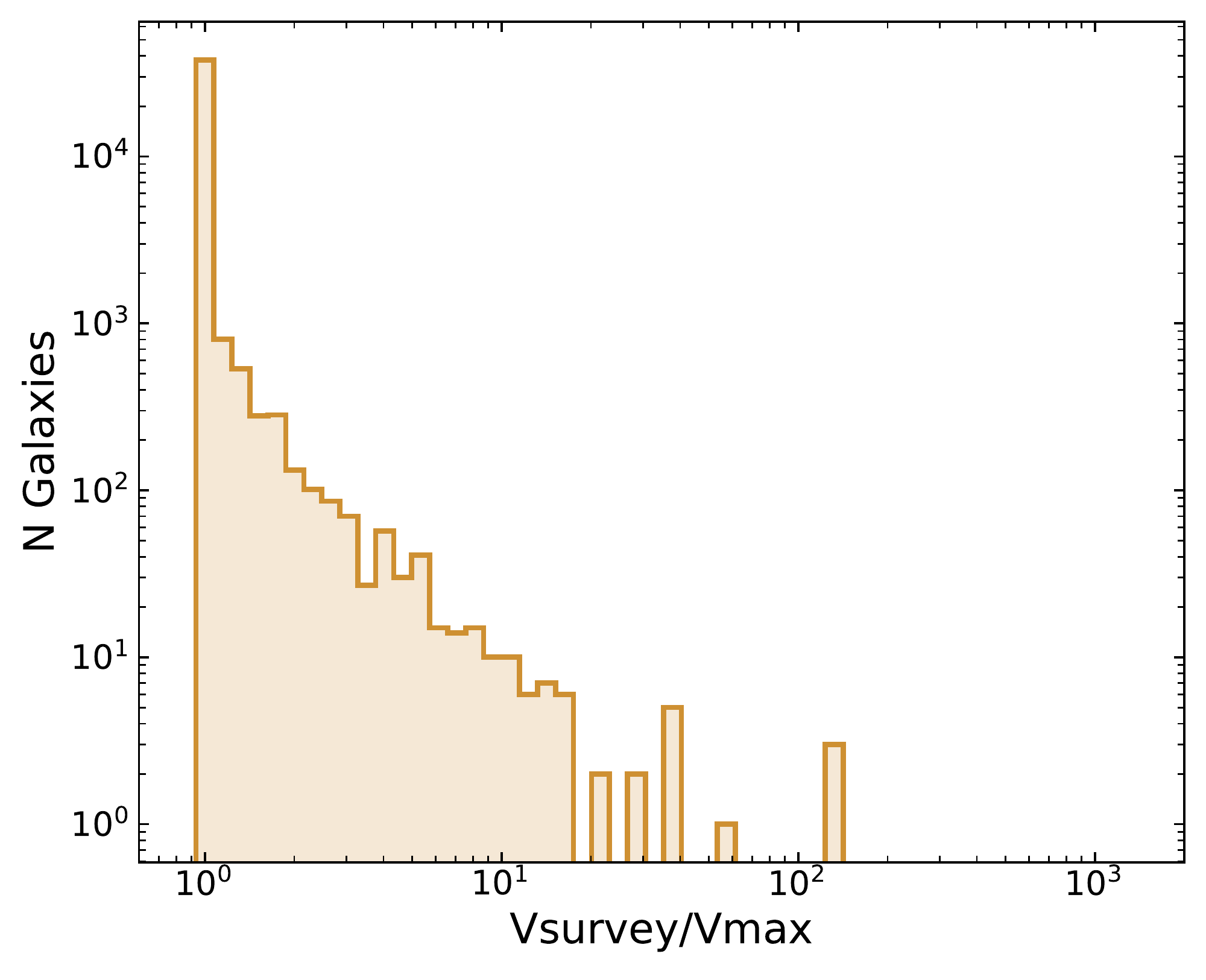}
    \caption{Distribution of $V_{\mathrm{max}}$ corrections applied to field galaxies in the DES sample in order to correct for incompleteness. The majority of galaxies have no correction, and the distribution of those that do follows a power-law.
    \label{fig:vmax_field}}
\end{figure}

\subsection{Quality cuts \label{subsec:cuts}}

In order to refine the sample of host and field galaxies for the rate analysis we perform a series of quality cuts:

\begin{enumerate}
    \item objects must be detected and have a Kron magnitude measurement in all four DES optical bands;
    
    \item field galaxies are limited to unmasked region of the SN-X3 field \citep{Hartley2020} which has an area of 1.52 deg$^2$;
    
    \item objects must not be within a given number (20 in $x$, 50 in $y$) of pixels of the CCD edge, as the co-addition of slightly misaligned images introduced a region of significant noise in this part of the detector;
    
    \item to minimise stellar contamination, objects must have a value of $<0.95$ in at least one band for the star/galaxy (S/G) separation metric \texttt{CLASS\_STAR} provided by \texttt{Source Extractor} \citep{Bertin1996};
    
    \item field galaxies must be covered by $ugrizJHK$ photometry, with a photometric flux measurement or upper limit present in all bands;
    
    \item objects must have a spectroscopic redshift measurement or a photometric redshift estimate with a well-defined peak in redshift  probability density;
    
    \item objects must be detected at signal-to-noise ratio (SNR) greater than 3 in the $r$ band;
    
    \item galaxies must be brighter than $m_r \leq 24.5$, as this is the magnitude of the faintest SN host with a spectroscopic redshift;
    
    \item galaxies must be within the redshift range $0.2 \leq z \leq 0.6$ (see Section \ref{subsec:incompleteness_SNe} for an explanation of this cut).
\end{enumerate}

The numbers of SN hosts and field galaxies passing these cuts are listed in Table \ref{tab:cuts}. The final samples comprise 809 SNe and their host galaxies and 40,415 field galaxies. The volume-weighted mean redshifts are 0.50 for both SNe and field galaxies.

\subsection{Galaxy properties \label{subsec:properties}}

We estimate global galaxy properties for both the SN host and field galaxies that pass the quality cuts by fitting the photometry with stellar population templates in a method outlined by \citet{Sullivan2006} and consistent to that used in previous DES-SN analyses \citep{Smith2020,Wiseman2020,Kelsey2021}. For SN hosts the redshift is fixed at the spectroscopically determined value, while for field galaxies we fix it at either the spectroscopic value if one exists in the OzDES GRC, or more commonly the photometrically derived value. We use the stellar population templates of \citet{Bruzual2003} and adopt a \citet{Chabrier2003} initial mass function (IMF). The fitting procedure returns a best fitting template and corresponding stellar mass ($M_*$). Upper and lower bounds on stellar mass estimates are taken as the extreme values that correspond to templates that are consistent with the data (given the photometric uncertainties) according to their $\chi^2$ statistic, as per \citet{Sullivan2006}. To check for bias caused by template choices, we also fit the galaxies using the P\'EGASE.2 templates \citep{Fioc1997,LeBorgne2002} and a \citet{Kroupa2001} IMF, and find results consistent within measurement uncertainties. This is consistent with the findings of \citet{Smith2020}. 

The results of our SED fitting are shown in Fig. \ref{fig:mass_hist_hosts_field}. The figure showcases the vastly different distributions of the two samples. SN hosts are preferentially high mass galaxies, whereas the field galaxy distribution increases down to lower masses, peaking around $10^{8.5}~\mathrm{M}_{\odot}$. The SN host stellar mass distribution is plotted twice: once as raw counts; once weighted by the host galaxy spectroscopic efficiency (Section \ref{subsec:incompletenss_SN_hosts}). As shown in Fig. \ref{fig:mass_hist_hosts_field}, the host stellar mass distribution of the DES-SN sample used in this analysis is qualitatively similar to that from the low redshift Palomar Transient Factory (PTF) sample \citep{Pan2014} as well as the large cosmological Pantheon sample \citep{Scolnic2018}. This consistency reflects that seen in the smaller spectroscopically confirmed samples presented in \citet{Wiseman2020} and \citet{Smith2020}. The field galaxy stellar mass distribution is shown on a log scale, along with the stellar mass function (SMF) from the ZFOURGE survey \citep{Tomczak2014} scaled to match at $\log (M/\mathrm{M}_{\odot}) = 10$, whose stellar masses are derived with the same photo-$z$ code, IMF and template library as used here. Above $10^{8.5}~\mathrm{M}_{\odot}$, the DES field galaxy distribution closely follows the ZFOURGE SMF for $0.5<z<0.75$, indicating that the DES sample is representative of field galaxies in this mass and redshift range.

\section{Incompleteness corrections}
\label{sec:incompleteness}
The simple ratio of the number of SN hosts and field galaxies presented at the end of the previous section provides a first approximation of the SN rate per galaxy, and the inclusion of a factor equal to the survey duration normalises the rate to per year. However, both the SN and field galaxy samples introduced in Sections \ref{subsec:host_sample} and \ref{subsec:field_sample} are affected by incompleteness, which is likely to be the dominant systematic effect in the analysis. In this section we describe our method of correcting for various sources of incompleteness in the data.

\subsection{Supernovae \label{subsec:incompleteness_SNe}}

Incompleteness in a SN survey arises from a number of sources. The primary source of incompleteness is caused by the magnitude limit of the survey: SNe with apparent magnitudes below the survey limit will not be detected. Since SNe Ia are relatively uniform in absolute luminosity, this form of incompleteness is primarily redshift dependent. The large redshift range of DES-SN also means that SNe are probed at different regions of their rest frame SED that vary significantly in luminosity. Additionally, DES-SN comprises 10 separate pointings, each with different visibility and thus airmass throughout the observing season. These differences lead to different detection efficiencies across the fields.

To correct for these incompleteness, we follow a similar method to that used in the PTF rates analyses of \citet{Frohmaier2019} and \citet{Frohmaier2020}, outlined in \citet{Frohmaier2017}. We simulate $1.1\times 10^6$ SNe in the redshift range $0.05 \leq z_{\mathrm{SN}} \leq 1.3$. The SNe are generated in the same way as for the training of SNN (Section \ref{subsubsec:sn_classify}). The SNe are simulated with explosion epochs $t_0$ uniformly distributed between two months before DES-SN began and two months after it finished in order to account for all SNe that could have been observed by DES-SN. We run mock versions of the DES-SN survey, using the exact cadence, conditions, and zeropoints from the survey itself. All of the detected simulated SNe are passed through the light curve fit of SALT2 \citep{Betoule2014}, as per the implementation in \texttt{SNANA}, and those that fail the light curve cuts outlined in Section \ref{subsubsec:sn_classify} are discarded. We are left with a fraction of the original simulated SNe, and that fraction is dependent on a combination of sky location, explosion epoch, redshift, stretch, and colour. The fraction of recovered SNe (the efficiency) is thus described by a 5-dimensional surface. For the $i$th SN the efficiency $\eta_{\mathrm{SN}, i}$ in field $F$, exploding at time $t_0$ at redshift $z$, with stretch $x_1$ and colour $c$, is:
\begin{equation}
    \eta_{\mathrm{SN},i} (F_i,z_i,t_{0,i},x_{1,i},c_i) = \left( \frac{N_{\mathrm{obs}}\left(F_i,z_i,t_{0,i},x_{1,i},c_i\right)}{N_{\mathrm{sim}}\left(F_i,z_i,t_{0,i},x_{1,i},c_i\right)}\right)\,.
\end{equation}

In practice, the gradient of the efficiency function is strongest between different DES fields and as a function of redshift, while SN stretch and colour have smaller effects. We integrate the efficiencies across the full simulated time range, and as such the efficiency is limited to $\sim 0.5$ due to the 6-month nature of the DES observing seasons.
The distribution of efficiency as a function of redshift is shown in Figure \ref{fig:SN_efficiency}. It is evident that the deep fields (X3, C3) are sensitive to SNe at higher redshifts, while there is no drastic shifts between efficiencies in the eight shallow fields. There is evidence that the SN efficiency depends weakly on $x_1$ and more strongly on $c$, which is expected since the colour correction term is larger than the stretch correction term in the SN Ia standardisation formula \citep{Tripp1998}. Blue SNe are recovered more readily than red SNe as they are generally brighter. 

The deep fields show non-zero efficiencies approaching $z=1$, whereas the shallow fields typically reach $z=0.8$. Since fractional uncertainty is large at such low efficiencies, we choose to make a redshift cut of $z=0.6$ where the efficiency is well above 0.1 in all fields. 

\subsection{Supernova hosts \label{subsec:incompletenss_SN_hosts}}
A further limiting factor in the SN host sample is the requirement of a spectroscopic redshift. The majority of SN host spectroscopic redshifts in DES are provided by the dedicated follow-up survey OzDES, for which the limiting magnitude is around 24 to 24.5 mag in the $r$ band \citep{Lidman2020}. The rate at which a host redshift is successfully measured given an apparent magnitude has been extensively modelled by \citet{Vincenzi2020} who provide the spectroscopic redshift efficiency $\epsilon_{z_{\mathrm{spec}}} (m^{\mathrm{host}}_{r,i})$ as a function of host $r$-band magnitude, host galaxy colour, and the year in which the SN was discovered in order to allow for a longer possible spectroscopic exposure time for hosts of SNe discovered earlier in the survey.  To assess how $\epsilon_{z_{\mathrm{spec}}} (m^{\mathrm{host}}_{r,i})$ translates into an efficiency as a function of redshift $\epsilon_{z_{\mathrm{spec}}} (z)$ within the redshift limits of our survey, we use a simulation of host galaxies as described in \citet{Vincenzi2020}. The simulation uses galaxy photometry from DES Science Verification images, and then selects them as potential SN hosts according a weighting driven by previously measured SN rate - host galaxy relationships. We note that this is for display purposes only, so the relationships built into the simulation are not propagated to the final rate measurements. For bins in redshift, we find the mean efficiency for measuring a spec-$z$ of all galaxies in the bin via $\epsilon_{z_{\mathrm{spec}}} (m^{\mathrm{host}}_{r,i})$. The resulting $\epsilon_{z_{\mathrm{spec}}} (z)$ curves are displayed in Fig. \ref{fig:eff_z} and show a high spectroscopic efficiency within the redshift limits of this analysis.

The effects of the host galaxy spectroscopic efficiency are displayed in the upper panel of Fig. \ref{fig:mass_hist_hosts_field}, where the normalised histogram weighted by the efficiency is skewed towards lower masses when compared to the unweighted distribution. So that we reduce any bias towards SNe in bright hosts that are easier to obtain spectroscopic redshifts for, the host galaxy spectroscopic redshift efficiency is multiplied by that derived from the SN detection efficiency (Section \ref{subsec:incompleteness_SNe}) to arrive at the final efficiency for the $i$th SN:
\begin{equation}
    \eta_{\mathrm{SN},i} = \eta_{\mathrm{SN},i} (F_i,z_i,t_{0,i},x_{1,i},c_i)\times \epsilon_{z_{\mathrm{spec}}} (m^{\mathrm{host}}_{r,i})\,.
\end{equation}

\subsection{Field galaxies \label{subsec:incompleteness_field}}

\subsubsection{Apparent magnitude limits \label{subsubsec:mag_lims}}
As we use photometric, rather than spectroscopic, redshifts for the field galaxies, they do not suffer from spectroscopic incompleteness as the SN hosts do. Instead, the inclusion of any given galaxy in the survey area in the sample is determined simply by whether it is detected above a prescribed threshold in signal-to-noise ratio, i.e. the sample is magnitude limited modulo the cuts described in Section \ref{subsec:cuts}. To determine the apparent magnitude limit for galaxies in each of the optical bands we employ the method of \citet{Johnston2007}, \citet{Teodoro2010}, and \citet{Johnston2012} (hereafter Completeness I, II, III respectively). These works provide two complimentary statistics in order to find the limiting magnitude based on ranking of galaxies' absolute magnitudes in a given band $M$ and distance moduli $Z$, named  $T_{\mathrm{C}}$ and $T_{\mathrm{V}}$, respectively. Galaxies are first placed in the $M-Z$ plane, which is limited to distance moduli that correspond to the redshift cuts $0.2 <z<0.6$ introduced in Section \ref{subsec:incompleteness_SNe}. Each galaxy in the survey is ranked by its $Z$ ($M$) compared to all other galaxies that fall within a slice of width $\delta M$ ($\delta Z$) -- we find that $\delta M = \delta Z = 0.02$~mag provides the best balance between high sampling resolution and an adequate number of galaxies within each slice. We then iteratively test different trial limiting apparent magnitudes $m_{\mathrm{lim, trial}}$. For a galaxy of apparent magnitude $m < m_{\mathrm{lim, trial}}$ observed in a survey complete to magnitude $m_{\mathrm{lim, true}}$ where  $m_{\mathrm{lim, trial}} \leq m_{\mathrm{lim, true}}$, the expectation value of the rank for a random is 0.5. However, if a trial limiting magnitude $m_{\mathrm{lim, trial}} \geq m_{\mathrm{lim, true}}$, there will be a lack of observed faint objects, such that the expectation value of the rank drops. 

Completeness I, II, and III show that for surveys with sharp, well defined magnitude limits the test statistics $T_{\mathrm{C}}$ and $T_{\mathrm{V}}$ are stable and flat as a function of apparent magnitude up until the magnitude limit, where they drop sharply. Fig. \ref{fig:completeness} shows the values of $T_{\mathrm{C}}$ and $T_{\mathrm{V}}$ in the DES X3 field. The values of the test statistics increase with $m_{\mathrm{lim, trial}}$ until a peak, before decreasing to stable values at magnitudes far beyond the limit of the survey, contrary to the sharp drop in an ideal survey. The shape at brighter $m_{\mathrm{lim, trial}}$ is likely caused by incompleteness at the bright end: due to the small sky area, we simply do not probe enough volume to sample the bright end of the galaxy luminosity function well enough for the statistics to be robust. 

To approximate an efficiency function, we fit the peak of the statistics with a polynomial function, and then normalise by the maximum and minimum values. We then interpolate between the peak and the faint-magnitude floor to find the 50\% completeness limit at the point where the normalised value of the test statistic is 0.5. These values are consistent with the value at which 50\% of the true sources are detected but have the advantage of being derived from the data without a simulation that is based on assumptions and thus susceptible to bias. Our final limiting magnitude for each band is the mean of that found from each of  $T_{\mathrm{C}}$ and $T_{\mathrm{V}}$.
\subsubsection{$V_{\mathrm{max}}$ correction \label{subsubsec:vmax_corr}}

To correct for incompleteness caused by the magnitude limited nature of the survey we follow the prescription of \citet{Sullivan2006} and \citet{Smith2012} by using a $V_{\mathrm{max}}$ method based on \citet{Schmidt1968}. For each galaxy in the sample we calculate the maximum volume within which it would have been observed given its absolute magnitude and $k-$correction, and the apparent magnitude limits calculated using the $T_{\mathrm{C}}$ and $T_{\mathrm{V}}$ statistics. A correction of $V_{\mathrm{survey}}/V_{\mathrm{max}}$ is applied to all galaxies for which $V_{\mathrm{max}} < V_{\mathrm{survey}}$, where $V_{\mathrm{survey}}$ is the maximum volume reached by the survey. In our case this corresponds to the volume at $z=0.6$. Fig \ref{fig:vmax_field} shows the distribution of the correction among galaxies in the sample. The vast majority of galaxies require no correction, meaning they would have been observed beyond the maximum volume considered. Roughly 1\% of objects have a correction greater than 1, with the distribution well described by a power-law function. The effect of the $V_{\mathrm{max}}$ correction on the total galaxy counts in each stellar mass bin can be seen in Fig. \ref{fig:mass_hist_hosts_field}; indeed, the two histograms are visibly indistinguishable above $10^9~\mathrm{M}_{\odot}$ which indicates a sample that is effectively complete in this mass range in the redshift range of interest.


\begin{figure}
    \centering
    \includegraphics[width=.5\textwidth]{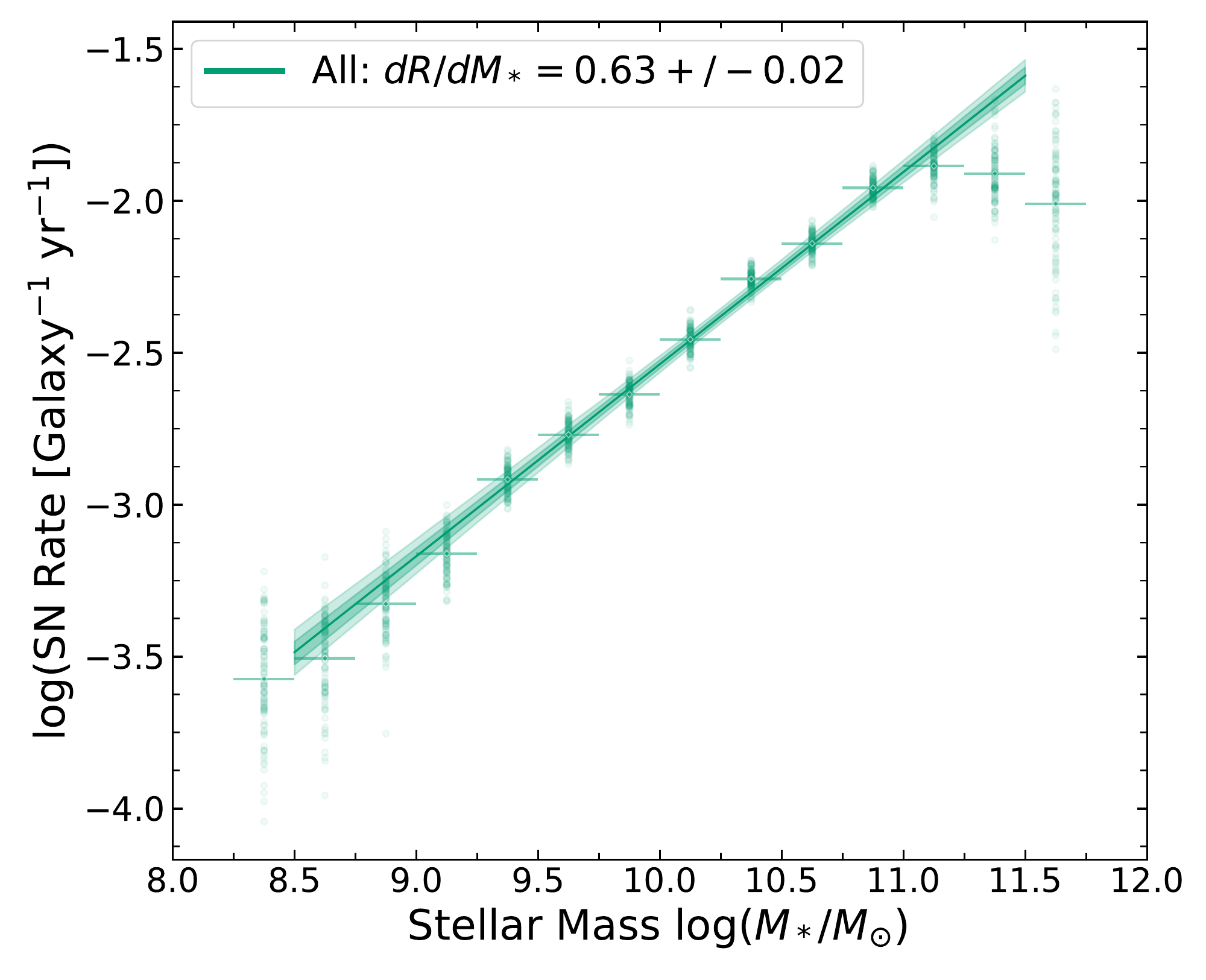}
    \caption{The rate per galaxy of SNe Ia as a function of stellar mass. Horizontal error bars represent the width of the stellar mass bins. Vertical error bars are estimated via a Monte-Carlo resampling of the rate given the uncertainties in stellar mass. The linear fit is based on all but the lowest and highest mass bins, and takes into account uncertainties in the rate.}
    \label{fig:rate_raw}
\end{figure}


\begin{figure*}
    \centering
    \includegraphics[width=\textwidth]{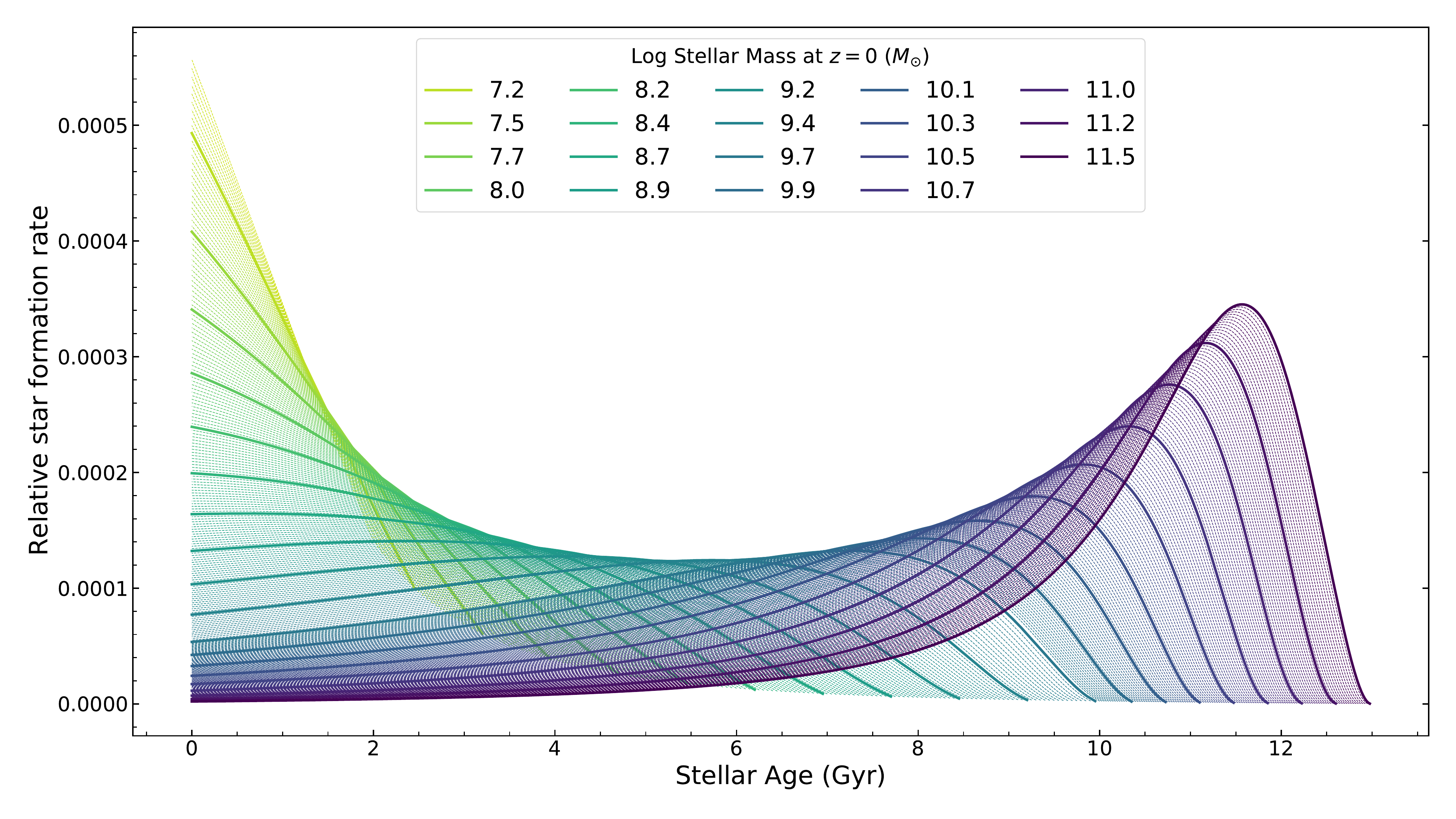}
    \caption{The stellar mass assembly of the Universe as prescribed by the toy model of \citetalias{Childress2014}. The $y$-axis is the fraction of a galaxy's total mass of formed stars that are formed in each time bin. Galaxies with high stellar mass at $z=0$ formed the majority of their stars in the distant past, while lower-mass galaxies are still actively star forming in the present day.}
    \label{fig:SFHs}
\end{figure*}

\begin{figure}
    \centering
    \includegraphics[width=.5\textwidth]{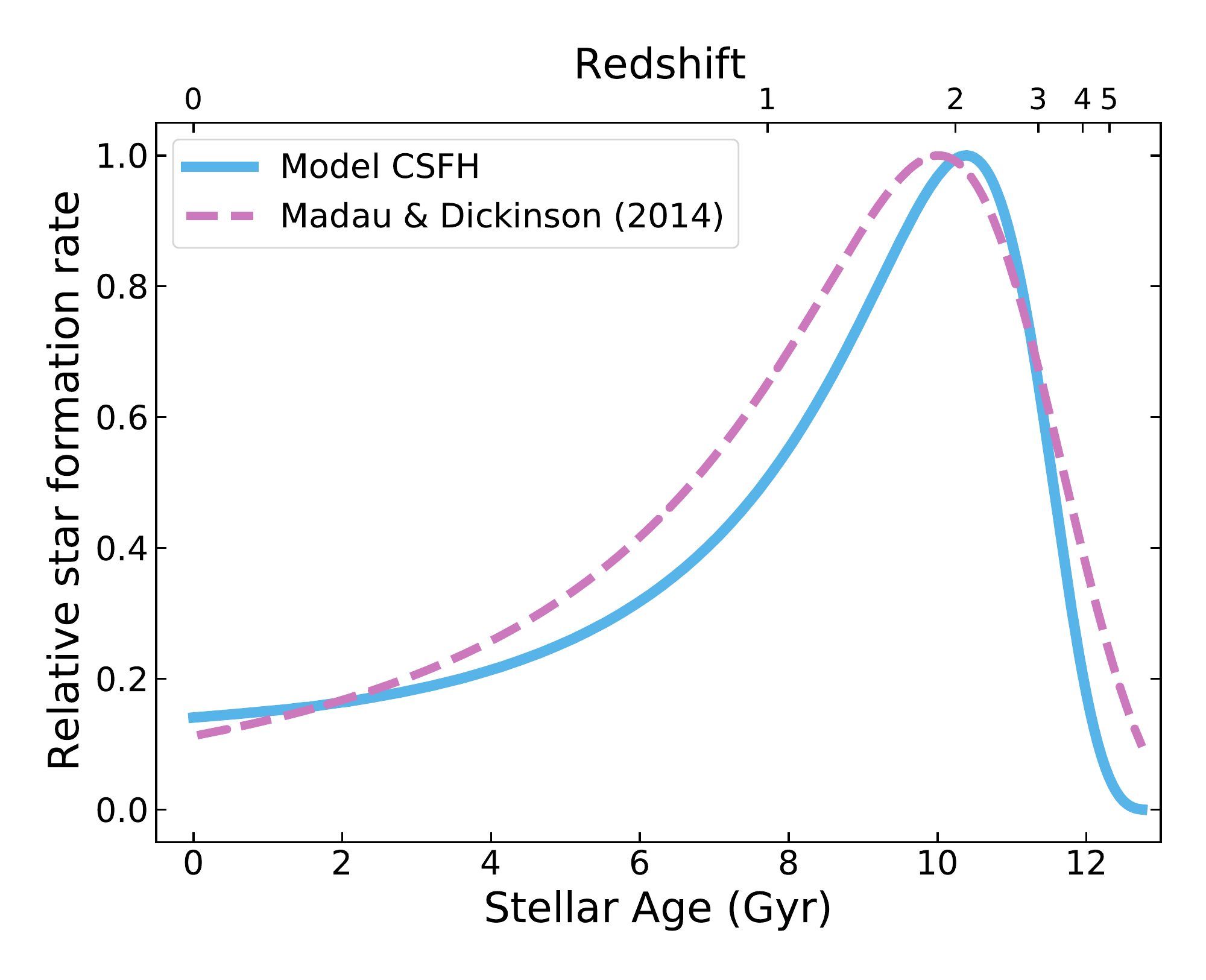}
    \caption{Cosmic star-formation history (CSFH) as approximated by our model (cyan solid) and as a best fit to observations (magenta dashed; \citealt{Madau2014}).}
    \label{fig:csfh}
\end{figure}


\section{The per-galaxy rate of type Ia supernovae}
\label{sec:rates}
The rate of SNe per galaxy ($R_{\mathrm{G}}$) per year in a transient survey can be approximated by the equation:
\begin{equation}
    R_{\mathrm{G}} = \frac{N_{\mathrm{SN}}}{N_{\mathrm{G}}} \frac{V_{\mathrm{G}}}{V_{\mathrm{SN}}}\frac{1}{T}\,,
\label{eq:rate1}
\end{equation}
where $N_{\mathrm{SN}}$ and $N_{\mathrm{G}}$ are the respective numbers of SNe and galaxies detected, $V_{\mathrm{SN}}$ and $V_{\mathrm{G}}$ the volumes from which the SNe and galaxy samples were taken, and $T$ the duration of the SN survey in years. 

As described in Section \ref{sec:incompleteness}, the values of $N_{\mathrm{SN}}$ and $N_{\mathrm{G}}$ that we observe are underestimates of the true numbers due to observational incompleteness -- we do not detect and count all SNe in the volume $V_{\mathrm{SN}}$, nor do we detect and count all galaxies in the volume $V_{\mathrm{G}}$. We thus estimate the intrinsic numbers of SNe and galaxies by multiplying the observed numbers by their respective incompleteness corrections calculated in Sections \ref{subsec:incompleteness_SNe} and \ref{subsec:incompletenss_SN_hosts} for SNe and Section \ref{subsec:incompleteness_field} for field galaxies:

\begin{equation}
    N_{\mathrm{SN,intrinsic}} = \sum_i^{n_{\mathrm{SN}}}\eta\left(F,z,t_0,x_1,c,m^{\mathrm{host}}_{r}\right)_{i}\,,
    \label{eq:corr_SN}
\end{equation}
and
\begin{equation}
    N_{\mathrm{G,intrinsic}} = \sum_j^{n_{\mathrm{G}}} \eta\left(V_{\mathrm{max}}\right)_{j} 
    \label{eq:corr_G}
\end{equation}
for each SN $i$ and galaxy $j$. The volumes $V_{\mathrm{SN}}$ and $V_{\mathrm{G}}$ are calculated from the sky areas from which the respective samples were taken: 1.52 deg$^2$ in the case of the field galaxies \citep{Hartley2020}; 23 deg$^2$ for SNe \citep{Smith2020a}. These areas are combined with the redshift interval $[0.2,0.6]$ to determine the total volumes.

\begin{figure}
    \centering
    \includegraphics[width=.5\textwidth]{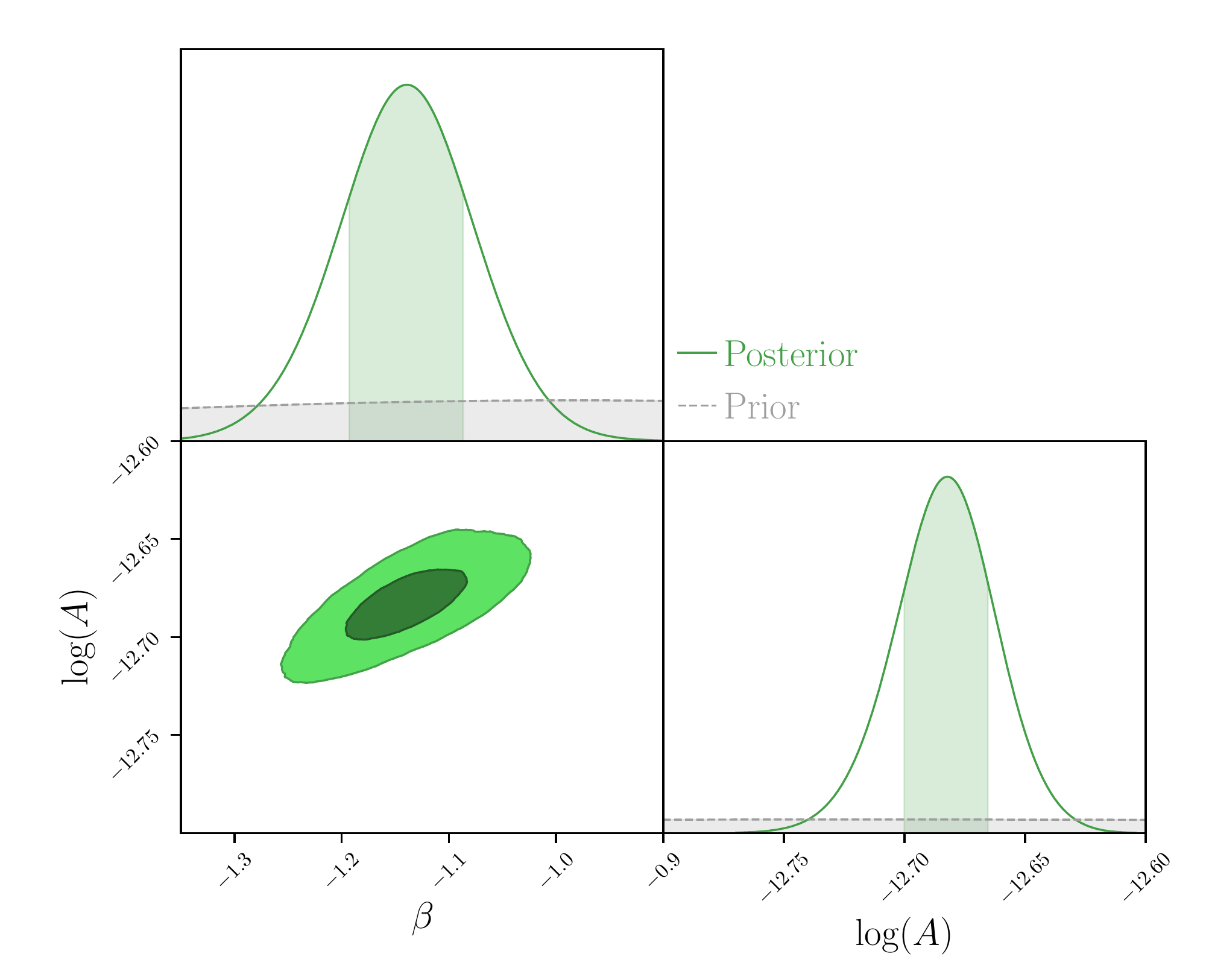}
    \caption{Posterior distributions for DTD slope $\beta$ and normalisation $A$. Displayed point estimates represent the median of the posterior samples.
    \label{fig:corner_beta_norm}}
\end{figure}
\begin{figure}
    \centering
    \includegraphics[width=.5\textwidth]{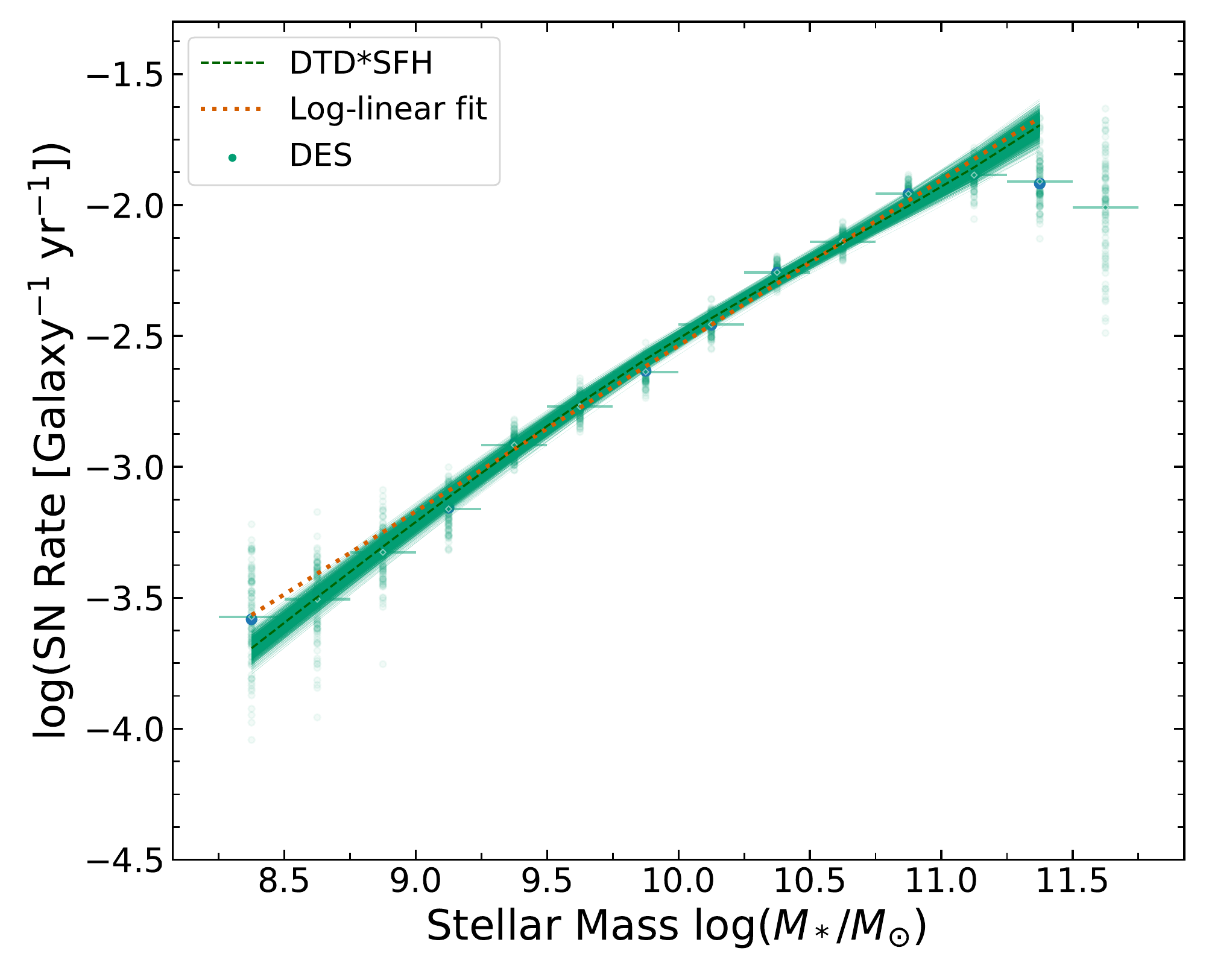}
    \caption{Rate per galaxy of SNe Ia as a function of stellar mass (green points) along with the prediction from the best fitting DTD parameters $\beta$, $A$, and $t_{\mathrm{p}}$ forward modelled through Eq. \ref{eq:galaxy_rate}. Samples from the posterior distribution of the model log rate are drawn in cyan while the black dashed line is the mean of the posterior. In red we show the simple log-linear fit from Section \ref{sec:rates}.
    \label{fig:rate_fitted}}
\end{figure}

\begin{figure}
    \centering
    \includegraphics[width=.5\textwidth]{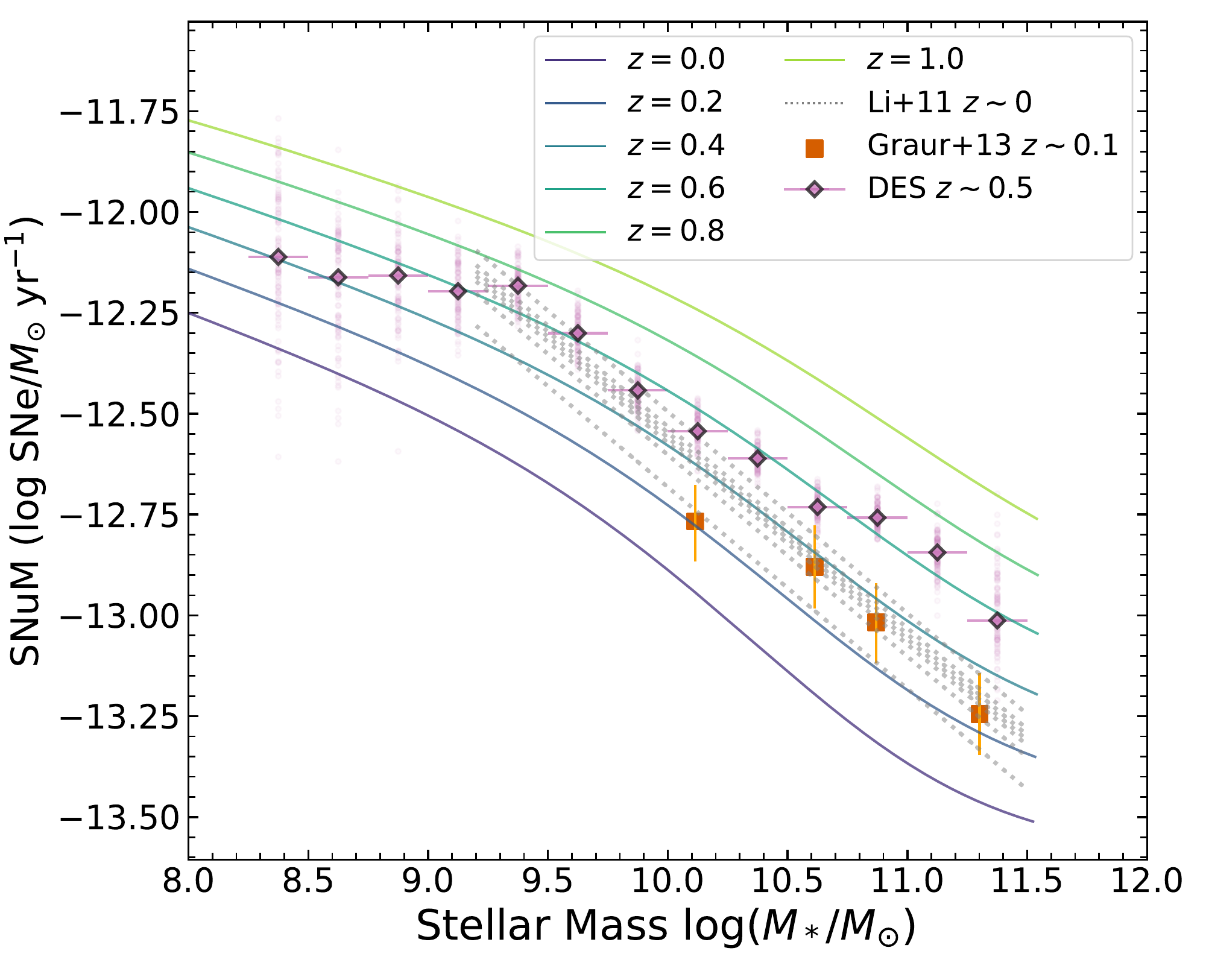}
    \caption{SN Ia rate per unit stellar mass as function of galaxy stellar mass. Error bars are the same as Fig. \ref{fig:rate_raw}. Literature data are that from SDSS spectra (orange squares; \citealt{Graur2013}) and the LOSS sample (grey dotted; \citealt{Li2011a}). The systematically higher SNuM in DES is caused by the higher redshift of the sample, as illustrated by the  prediction of the model using our best fit DTD from Section \ref{subsec:results_dtd} (coloured curves). }
    \label{fig:snum}
\end{figure}
By binning both the SN hosts and field galaxies by their stellar mass we measure the mean SN Ia rate per galaxy as a function of stellar mass, $R_{\mathrm{G}}(M_*)$. We employ a bootstrap Monte-Carlo approach in order to estimate the uncertainty in the value of the rate in each stellar mass bin. The probability density function (PDF) of each galaxy's stellar mass is represented by the sum of two half Gaussian distributions to represent the asymmetric positive and negative uncertainties derived in the SED fitting stage (Section \ref{sec:data}). For each SN host and field galaxy, we take 100 samples of its stellar mass by drawing at random from its PDF. For each of the 100 samples we calculate the SN rate per galaxy in each stellar mass bin via Eq \ref{eq:rate1}, using the completeness-corrected values of $N_{\mathrm{SN}}$ and $N_{\mathrm{G}}$ from Eqs. \ref{eq:corr_SN} and \ref{eq:corr_G} respectively. 

The rate of SNe Ia per galaxy as a function of stellar mass is shown in Fig. \ref{fig:rate_raw}. The relationship between SN Ia rate and galaxy stellar mass is well described by a linear function in log space, which corresponds to a power law. To find the slope and intercept that best describe the data we use Bayesian inference. To sample the posterior we use the enhanced no-U-turn Sampler (NUTS) algorithm \citep{Betancourt2017}, which is a variant of Hamiltonian Monte-Carlo (HMC), implemented in the Stan programming language \citep{Carpenter2017}. We describe the fitting procedure in full detail in Appendix \ref{appendix:linear_fits}. We measure a slope of $0.63\pm0.02$, which is consistent with the value found in weakly star-forming galaxies in the Supernova Legacy Survey by \citet{Sullivan2006}, but also in passive galaxies in SDSS by \citet{Smith2012}, and in all galaxies in the Lick Observatory Supernova Search (LOSS) by \citet{Li2011a}. Our slope is also consistent with that measured in the low-redshift All-Sky Automated Survey for Supernova (ASAS-SN) Bright Supernova Catalogues \citep{Brown2019}, which extends to galaxies with stellar mass as low as $10^7$~M$_{\odot}$.

A best fitting slope of less than unity corresponds to a power-law, and indicates that SN Ia rate is not uniquely determined by host galaxy stellar mass. Indeed, the rate is instead likely to be driven by the DTD which is a non-linear function of stellar age. Investigating the shape of this DTD forms the basis for the latter part of this paper in Sections \ref{sec:model} and \ref{sec:split_x1_c}. The straight line fit in Fig. \ref{fig:rate_raw} deviates somewhat from the central values of the data at high stellar mass, the reason for which we investigate in Section \ref{subsec:discussion}.

\section{Modelling the per-galaxy rate of type Ia supernovae}
 \label{sec:model}
 
In Section \ref{sec:rates} and Fig. \ref{fig:rate_raw} we showed that the SN Ia rate per galaxy as a function of stellar mass is approximated by, but not perfectly described by, a power-law. In this section, we introduce a physical model in order to better fit the observed rate vs stellar mass relationship. 

We begin by considering the delay time distribution of SNe Ia.
We represent the SN Ia DTD by a power law, which we consider effective after some "prompt time" $t_{\mathrm{p}}$, before which its value is set to 0. $t_{\mathrm{p}}$ is generally interpreted as the time taken for WDs to form after the burst of star formation. The DTD is thus:
\begin{equation}
 \Phi(\tau) = \left\{
    \begin{array}{@{}cc@{}}
    0 , & \tau < t_{\mathrm{p}} \\
    A\left(\frac{\tau}{\mathrm{Gyr}}\right)^{\beta} , & \tau \geq t_{\mathrm{p}}
    \end{array}\right.
        \label{eq:dtd}
\end{equation} 
where $\tau$ is the time since a burst of star formation, and $A$ is a normalisation at $\tau=1$ Gyr. If all stars in a galaxy were formed at a single epoch $t_f$, the DTD would describe the rate of SNe at an observation time $\tau = t_0$. However in reality galaxies are formed by the gradual build up of stellar mass over several epochs of star formation -- the distribution of this mass build up is known as the star-formation history (SFH). The rate of SNe Ia is thus the sum of the DTD evaluated for each epoch of star formation, multiplied by the stellar mass formed in that epoch. Mathematically this is represented by the convolution of the DTD and SFH:
 
\begin{equation}
    R_{\mathrm{G}} = \int_{t_0}^{t_f} \psi(t_0-\tau)\Phi(\tau)\,d\tau \,,
    \label{eq:galaxy_rate}
\end{equation}
where $t_0$ is the epoch at which the galaxy is observed, $t_f$ is the time at which the first stars in the galaxy formed, and $\psi$ is the SFH. 

In previous work \citep[e.g.][]{Strolger2004,Maoz2012} it has been common to determine the SFH for every galaxy in the survey via SED fitting. Eq. \ref{eq:galaxy_rate} is then used to calculate an expected number of SNe in each galaxy given the effective survey time, which is compared to the observed number in that galaxy (usually 0 and occasionally 1) using Poisson statistics. This method relies on either photometry covering several wavelength bands, or optical spectra with a high signal-to-noise ratio (S/N), in order to distinguish accurately between SFHs. Such accuracy is not possible for the DES sample; we do not possess spectra of all galaxies in the field, and in some cases only have the four optical bands available -- and have a maximum of eight from NUV to NIR -- from which to infer a SFH. This lack of detail is compounded by our reliance on photometric redshifts, which add an extra layer of uncertainty to the calculation. 

Instead of determining individual SFHs for each galaxy in our field sample, we use an empirical approach to estimate mean SFHs for galaxies as a function of their stellar mass at any given redshift (or equivalently, time $t_0$), that can be represented as $\hat \psi \left(t_0 -\tau; M_* \right)$. This method paves the way for modelling the SN Ia rate as a function of stellar mass by combining the mean SFHs and the DTD through Eq. \ref{eq:galaxy_rate}, and thus placing constraints on the DTD.

\subsection{Modelling the star formation histories of galaxies \label{subsec:method_sfh}}

To model the SFH of galaxies as a function of their stellar mass we adopt the prescription of stellar mass assembly of \citetalias{Childress2014}, which draws on the work of \citet{Zahid2012}. In the model, galaxies are expected to evolve smoothly along the so-called "main sequence of star formation" whereby the SFR of a galaxy of stellar mass $M_*$ at redshift $z$ is determined by a simple relationship (the SMz relation, \citealt{Zahid2012}). Specifically, we implement their "fine-tuned" model (Eq. A7 in \citetalias{Childress2014}), whereby the SMz flattens above $z\sim2$ in line with observations \citep{Stark2013}:

\begin{equation}
    \Psi(M_*,z) = \left(\frac{M_*}{10^{10}}\right)^{0.7}\frac{\exp{\left(1.9z\right)}}{\exp{\left(1.7\left(z-2\right)\right)} + \exp{\left(0.2\left(z-2\right)\right)} } [\mathrm{M}_{\odot} \mathrm{yr}^{-1}]\,,
    \label{eq:psi_mz}
\end{equation}
where $M_*$ is given in units of $\mathrm{M}_{\odot}$.
As galaxies grow, their SFR begins to slow down and eventually shut off almost entirely in a process known as quenching. We adopt the quenching penalty $p_Q$ directly from \citetalias{Childress2014}:
\begin{equation}
    p_Q(M_*,z) = \frac{1}{2}\left[1 - \mathrm{erf}\left(\frac{\log(M_*) - \log(M_Q(z))}{\sigma_Q}\right)\right]\,,
    \label{eq:pq}
\end{equation}
where $M_Q(z)$ describes how the quenching mass evolves with redshift, and $\sigma_Q$ is the transition scale which controls how fast a galaxy quenches. We adopt the observationally motivated form of the quenching mass evolution (Eq. A8 from \citetalias{Childress2014}):
\begin{equation}
    \log(M_q(z)/\mathrm{M}_{\odot}) = 10.077 + 0.636z \,, 
    \label{eq:mq}
\end{equation}
and a transition scale of $\sigma_Q = 1.1$ which provides a good fit to data from the Galaxy And Mass Assembly (GAMA) survey \citep{Baldry2012}.

Finally, at time $t$, some time $\tau$ since an epoch of star formation, a fraction of stellar mass is lost. We adopt the parametrization used by \citetalias{Childress2014}, namely that of \citet{Leitner2011} for a Chabrier 2003 IMF:

\begin{equation}
    f_{\mathrm{ml}} = 0.046\ln \left(\frac{\tau}{0.276~\mathrm{Myr}}+1 \right)
    \label{eq:fml}
\end{equation}
The mass loss at any given time is thus the convolution of the Eqs. \ref{eq:psi_mz} and \ref{eq:fml}:
\begin{equation}
    \Delta M_* = \int_0^{t_f} \Psi(M_*(t-\tau),z(t-\tau))\cdot\left(f_{\mathrm{ml}}(\tau + \Delta t) - f_{\mathrm{ml}}(\tau)\right)\mathrm{d}\tau\,.
    \label{eq:mlt}
\end{equation}
The stellar mass formed in each time $t$ (and corresponding redshift $z$) is thus the combination of the mass- and redshift-dependent SFR (Eq. \ref{eq:psi_mz}), reduced by the mass- and redshift-dependent quenching penalty (Eq. \ref{eq:pq}), minus the mass lost in each time step (Eq. \ref{eq:mlt}):

\begin{equation}
    \frac{M_*(t+\Delta t) - M_*(t)}{\Delta t} = p_Q\left(M_*(t_,z(t)\right)\cdot\Psi(M_*(t),z(t)) - \frac{\Delta M_*}{\Delta t}\,.
    \label{eq:sfh_model}
\end{equation}

As per \citetalias{Childress2014} we plant seed galaxies with initial masses of $10^6 ~\mathrm{M}_{\odot}$ at intervals of 50 Myr for look-back times $1 \leq t_f \leq 10$ Gyr, and 25 Myr for look-back times $10 \leq t_f \leq 13$ Gyr to achieved a more detailed model at early times with high SFR. For each seed galaxy, we evaluate Eq. \ref{eq:sfh_model} at time steps of 0.5 Myr and record the total mass formed and lost in each of the time steps -- this defines our model SFH to be used as input to Eq. \ref{eq:galaxy_rate}.

The result of our toy model is shown in Fig. \ref{fig:SFHs}. Galaxies with high masses at $z=0$ formed the vast majority of their stars in the first few Gyr and are now likely to be passive with old stellar populations, while low-mass galaxies are currently strongly star-forming with a vast majority of young stars. Galaxies with masses around $10^{10}~\mathrm{M}_{\odot}$ are composed of a mixture of young and old stellar populations. 
\subsubsection{Validating the mass assembly model \label{subsubsec:validate_c14}}

As noted in \citetalias{Childress2014}, this toy model by construction follows several empirical relations derived from galaxy observations: model galaxies follow the SMz relation and obey quenching prescriptions motivated by observed galaxies. Here we make two further checks to validate that the model can be used to approximate the stellar age distribution of SN Ia host galaxies in DES.

Firstly, we follow \citetalias{Childress2014} by summing our SFHs and weighting by the galaxy stellar mass function at $z=0$ to estimate the CSFH. We approximate the present-day stellar mass function according to a double Schechter function with parameters from the Galaxy and Mass Assembly survey \citet{Baldry2018}: $\log(M^*/\mathrm{M}_{\odot})=10.66$, $\phi^*_1=3.96\times10^{-3}$, $\alpha_1=-0.35$, $\phi^*_2=0.79\times10^{-3}$, $\alpha_2=-1.47$.
The comparison of our approximate CSFH with that measured from observations \citep{Madau2014} is shown in Fig. \ref{fig:csfh}. As was found in \citetalias{Childress2014}, the two functions show a good qualitative agreement -- in fact, we find a closer match between the model and observed CSFH, accurately reproducing the peak of cosmic star formation around $z\sim2$ and tracing the declining SFR to the present day.

Secondly, we assess the shape of our SFHs and how they depend on stellar mass by comparing with SFH measurements from the Calar Alto Legacy Integral Field spectroscopy Area survey (CALIFA) as revealed by \citet{GonzalezDelgado2017}. Our models agree qualitatively on numerous features of the SFHs (their Fig. 5): high-mass galaxies form almost all their stars at high redshift; intermediate-mass galaxies display a plateau in their SFR from $\sim4$~Gyr to the present day; low-mass galaxies show an increasing rate of star formation at $z=0$. The main discrepancy with the lowest-mass bin of the CALIFA sample $8.6 <\log (M_*/\mathrm{M}_{\odot})<9.8$, which as well as an increasing SFR at low redshift shows a second peak of star formation at large lookback times, which we do not see in our model. However, this mass bin spans over one order of magnitude in stellar mass. Upon closer inspection of our model, the lower edge of this mass bin corresponds to galaxies that form at a lookback time of 5~Gyr and display a rising SFH. At the upper edge, galaxies do indeed form at times $>10$~Gyr and display an early peak followed by exponential decline. The combination of these different model SFHs in the low-mass bin thus explain the dual-peaked nature of the observed SFHs in CALIFA.

\renewcommand{\arraystretch}{1.2}
\begin{table}
	\centering
	\caption{Results of the Bayesian parameter estimation for the SN Ia DTD.}
	\label{tab:dtd_results}
	\begin{tabular}{lccr} 
		\hline
		 &$\beta$ & $A$ & $t_{\mathrm{p}}$\\
		 &-       & $10^{-13}$ M$_{\odot}^{-1}$yr$^{-1}$ & Gyr \\
		\hline
		Fixed $t_{\mathrm{p}}$ & $-1.13^{+0.04}_{-0.06}$ &  $2.11^{+0.05}_{-0.12}$ & -\\
		Fixed $A$, $\beta$ & - & - & $0.047_{-0.007}^{+0.008}$\\
		\hline
	\end{tabular}
\end{table}

\subsection{Constraints on the SN Ia delay time distribution}
\label{subsec:results_dtd}

At the beginning of this Section, we showed how the rate of SNe Ia is driven by the convolution of the SFH of galaxies and the SN Ia DTD (Eq. \ref{eq:galaxy_rate}). We then prescribed a model to infer the mean SFHs of galaxies of any given stellar mass. Here, we measure the DTD by forward modelling it through Eq. \ref{eq:galaxy_rate} at the stellar masses corresponding to the centres of the bins in Fig. \ref{fig:rate_raw}. 

We assume a DTD $\Phi(\tau)$ with a power-law form described by index $\beta$ (Eq. \ref{eq:dtd}), normalisation $A$ and effective after prompt time $t_{\mathrm{p}}$. As per Section \ref{sec:rates}, we constrain the parameters via Bayesian inference, using HMC to explore the posterior distribution. At each step of the sampling procedure the model is re-calculated via Eq. \ref{eq:galaxy_rate}, we evaluate the likelihood assuming that the rate measurements $R_{\mathrm{G}, M_*}$ are described by a Gaussian PDF with mean $\hat R_{\mathrm{G}, M_*}$ and standard deviation $\sigma_{R_{\mathrm{G}, M_*}}$. For $\beta$ we adopt a Gaussian prior with hyper-parameters $p(\beta) \sim \mathcal{N}(-1\,,0.5)$. We fit for $A$ in log space, and adopt a Gaussian prior (in log space) with hyper-parameters $p(\log(A)) \sim \mathcal{N}(-12.7\,, 0.5)$.

SNe Ia do not occur instantaneously after an episode of star formation. Instead stars with zero-age main-sequence (ZAMS) masses less than $8~\mathrm{M}_{\odot}$ take time to evolve along the main sequence and form WDs and this time is not well constrained. In many works this ``prompt time" which we denote $t_{\mathrm{p}}$ is fixed to some value expected to be the minimum time for a star to evolve off the main sequence and become a WD, such as 40~Myr \citep{Maoz2012,Graur2013,Graur2014}. Other studies have left $t_{\mathrm{p}}$ as a free parameter \citep{Heringer2019,Castrillo2020}. We perform three fits: one with $t_{\mathrm{p}}$ fixed at 40 Myr, one with it as a free parameter along with $\beta$ and $A$, and a third with $\beta$ and $A$ fixed and  $t_{\mathrm{p}}$ free. We fit $t_{\mathrm{p}}$ in log space and adopt a Gaussian prior with hyper-parameters $p(\log(t_{\mathrm{p}})) \sim \mathcal{N}(-1.3\,,0.5)$, which corresponds to being centred around 50~Myr. Choosing this regime to fit allows more prior weight to be placed on shorter prompt times as per the majority of the literature. We use the $\hat{R}$ diagnostic of \citet{Vehtari2019} to assess the convergence of MC chains. When including $t_{\mathrm{p}}$ as a free parameter, we find that the strong degeneracy between $t_{\mathrm{p}}$ and $A$ cause the sampler to fail to converge with $\hat{R} \sim 1.3$ for the chains of those two parameters. In subsequent analyses we adopt as our fiducial model that with fixed $t_{\mathrm{p}} =0.04$~Gyr.

The joint and marginal posterior distributions for the fit parameters are found in Fig. \ref{fig:corner_beta_norm}, and the posterior means and standard deviations are summarised in Table \ref{tab:dtd_results}.  Fig. \ref{fig:corner_beta_norm} shows that there is a mild correlation between $\beta$ and $A$. In Fig. \ref{fig:rate_fitted} we show the SN rate per galaxy predicted by the model assuming the best fitting DTD parameters from Table \ref{tab:dtd_results} compared to the data as presented in Section \ref{sec:rates}. Visually, the model provides a good fit across a wide range of stellar mass. The model clearly diverges from the simple log-linear fit, describing better the enhanced rate around $10^{10}$ M$_{\odot}$ and suppression at high masses. This improvement provides evidence to support the DTD*SFH model. However, the model prediction still diverges from the data at high stellar masses ($>10^{11.25}$ M$_{\odot}$), suggesting that these data cannot be explained by our model, but are driven by further processes that we have not included. We discuss these in Section \ref{subsec:discussion}. 

\subsection{Comparison to previous DTD measurements \label{subsec:compare_dtd}}

\subsubsection{The DTD power law index \label{subsubsec:compare_beta}}
We measure a DTD power law index of $-1.13^{+0.04}_{-0.06}$. This value is consistent with the majority of previous analyses using various methods. Our measurement provides an independent support to the values close to $-1$ found using volumetric rates \citep[e.g.][]{Graur2011,Graur2014,Frohmaier2019}, individual galaxy SFHs \citep[e.g.][]{Maoz2012,Graur2013}, and galaxy clusters \citep[e.g.][]{Maoz2010} although our slope is marginally inconsistent with that of \citet{Heringer2019}. The result is also qualitatively consistent with that of \citet{Strolger2020a} who fit an exponential function DTD rather than a power law. However, their best fit model indicated a higher SN rate at long delay times than a $\tau^{-1}$ DTD, which is opposite to the mild preference of our fit which is steeper and thus produces a smaller fraction of SNe at long delay times.

\subsubsection{The DTD normalisation \label{subsubsec:compare_A}}
We measure a DTD normalisation (i.e. the rate of SNe 1 Gyr after star-formation) of $2.11^{+0.05}_{-0.12} \times 10^{-13}$~M$_{\odot}^{-1}$~yr$^{-1}$. \citet{Graur2013} found a value of $0.7 \times 10^{-13}$~M$_{\odot}^{-1}$~yr$^{-1}$ using the SFH technique, whereas \citet{Heringer2019} report $7 \times 10^{-13}$~M$_{\odot}^{-1}$~yr$^{-1}$. These results lie either side of our measured value. \citet{Heringer2019} note that the normalisation recovered from the SFHR method is degenerate with the assumed prompt time $t_{\mathrm{p}}$, a degeneracy that is also reflected in the non-convergence of our fit that included $t_{\mathrm{p}}$ as a free parameter. The normalisation is also sensitive to the assumed IMF which can explain this variation \citep{Maoz2017}. 

By integrating the normalised DTD over cosmic time, we obtain an estimate of the average SN Ia efficiency $N_{\mathrm{Ia}}/M_*$ which represents the number of SNe Ia formed per unit mass of stars formed. We measure $N_{\mathrm{Ia}}/M_* = 0.9~_{-0.7}^{+4.0} \times 10^{-3}~\mathrm{SNe}~\mathrm{M}_{\odot}^{-1}$, which is consistent with \citet{Graur2011}, \citet{Maoz2011}, \citet{Perrett2012}, and \citet{Graur2013}

\subsubsection{The SN Ia prompt time \label{subsubsec:compare_tp}}

The majority of works in the field have constrained the DTD power-law slope, and many also estimate its normalisation. Conversely, few have attempted to constrain the prompt time $t_{\mathrm{p}}$. In most cases, $t_{\mathrm{p}}$ has been fixed at some fiducial value such as 40~Myr, which is derived from the lifetime of 8~M$_{\odot}$ stars. \citet{Castrillo2020} included $t_{\mathrm{p}}$ (which they denote $\Delta$) in their fit, and find a value of $50_{-35}^{+100}$~Myr which is consistent with both 40~Myr. While we are unable to fit $t_{\mathrm{p}}$ simultaneously with $\beta$ and $A$, by fixing $\beta=-1$ and $\log(A)=-12.75$ we find $t_{\mathrm{p}} =47_{-7}^{+8}$~Myr, which is consistent with the fiducial value of $40$~Myr. \citet{Heringer2019} also present fits with varying prompt times, although they don't fit for the parameter itself. They find that varying $t_{\mathrm{p}}$ changes the recovered normalisation, which is consistent with the degeneracy encountered in our fits. Overall, we conclude that there is no strong evidence for a value of $t_{\mathrm{p}}$ that is different to the fiducial value of 40~Myr and adopt this value for the remainder of the analysis.

\begin{figure}
    \centering
    \includegraphics[width=.5\textwidth]{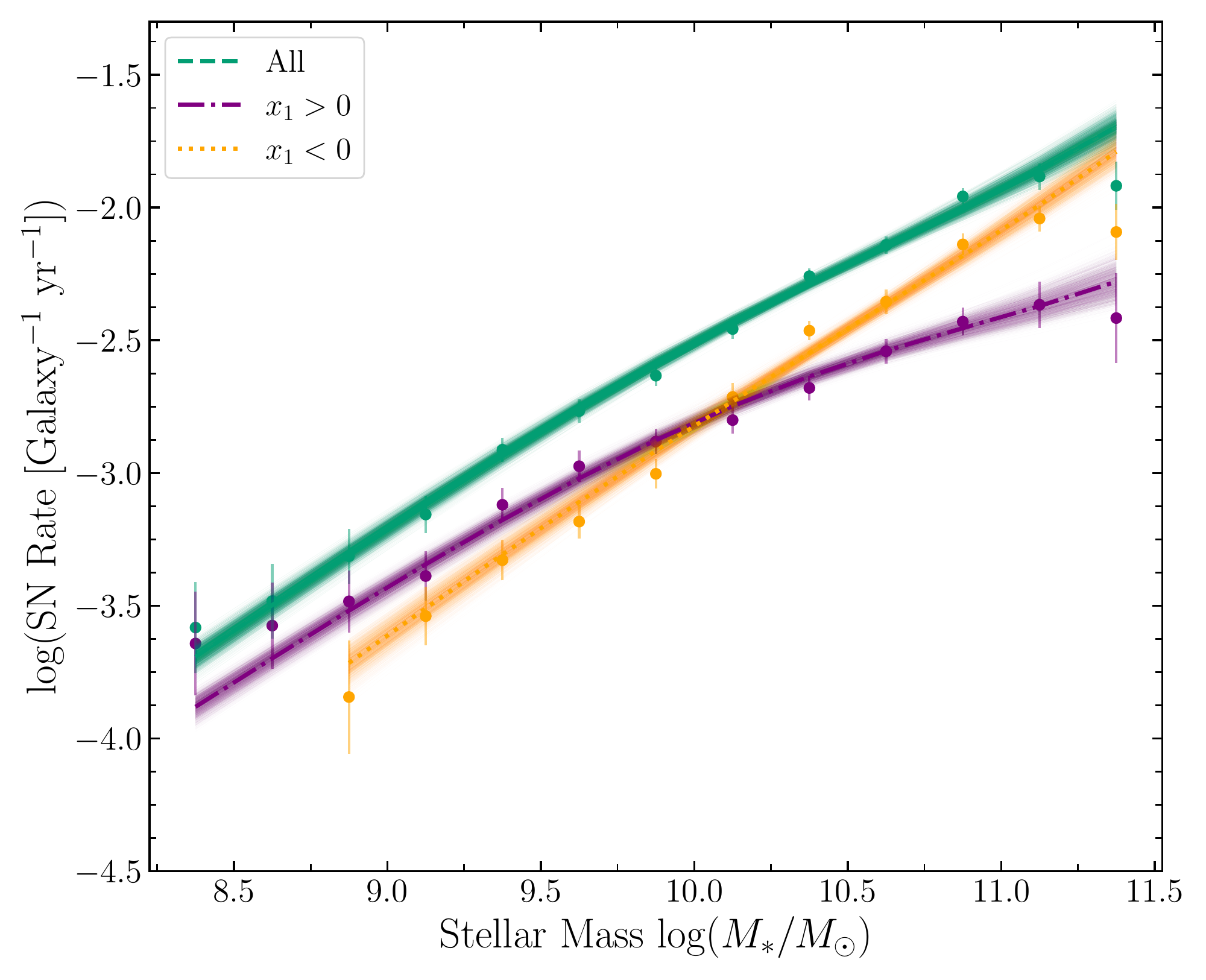}
    \caption{As per Fig. \ref{fig:rate_fitted} but with the data split by their stretch parameter $x_1$. Fits were performed with the DTD slope $\beta$ and normalisation $A$ as free parameters, and prompt time $t_{\mathrm{p}}$ fixed at 0.04 Gyr.
    \label{fig:rate_fitted_split_x1}}
\end{figure}

\renewcommand{\arraystretch}{1.2}
\begin{table}
	\centering
	\caption{Results of the Bayesian parameter estimation for the SN Ia DTD.}
	\label{tab:dtd_results_split_lc}
	\begin{tabular}{crr} 
		\hline
		 Sample &$\beta$ & $A$ \\
		 &-       & $10^{-13}$ M$_{\odot}^{-1}$yr$^{-1}$ \\
		\hline
		Fiducial & $-1.13\pm0.05$ &  $2.11\pm0.08$ \\
	    $x_1 < 0$ & $-0.79\pm0.08$ &  $1.19 \pm0.05$ \\
		$x_1 > 0$ & $-1.70^{+0.19}_{-0.10}$ & $0.50^{+0.26}_{-0.09}$ \\
		$c < 0$ & $-1.11^{+0.05}_{-0.11}$ & $0.90\pm0.04$ \\
		$c > 0$ & $-1.18\pm0.09$ & $1.14^{+0.11}_{-0.05}$ \\
		\hline
	\end{tabular}
\end{table}
\subsection{Second order processes affecting the supernova rate \label{subsec:discussion}}

The use of SN Ia rate measurements to constrain the SN Ia DTD has consistently led to a $t^{-1}$ power-law, which is widely accepted as evidence supporting the DD scenario. While our results are also consistent with those derived in previous studies, our model is mildly inconsistent with the SN rate per galaxy at very high stellar mass. The model over predicts the observed rate of SNe Ia in galaxies in that stellar mass range, caused either by a miscalculation of the rate or second order effects acting to suppress the rate compared to the fiducial model.

\subsubsection{Simplifications in the galaxy evolution model}

The toy model of stellar mass assembly used in this work includes several assumptions about the evolution of galaxies. In particular, it is assumed that galaxies evolve independent of each other, growing up the star formation main sequence until they reach a mass at which they quench and star formation ceases. While these simplifications describe the average properties of galaxies well (\citealt{Zahid2012},\citetalias{Childress2014}), they struggle to describe the more stochastic nature of galaxy evolution that occurs at the low and high mass ends: starbursts and quenching episodes, respectively. 

At some point along the evolutionary pathway, galaxies begin to cease star formation due to a combination of processes that together are known as quenching. In our model of mass assembly the characteristic mass at which quenching occurs is described by Eq. \ref{eq:mq}, and the rate of the transition from star forming to passive (as a function of the stellar mass) is determined by the transition scale $\sigma_{\mathrm{Q}}$ which we set to 1.1 based on GAMA observations. It is possible that this transition is too narrow and that quenching happens too fast in our model, leading to an under-prediction of the prompt fraction of SNe in the highest mass galaxies. To address this issue we rerun the mass assembly model with $\sigma_{\mathrm{Q}} = 1.5$ as per the nominal analysis of \citetalias{Childress2014} and refit the SN Ia rate data. We find that the adapted quenching model does not result in a better fit to the high-mass turnover in the SN rate.

\subsubsection{Effects of stellar metallicity}

One possible cause of the discrepant SN Ia rate at high stellar mass is the effect of stellar metallicity. Metallicity has been previously invoked to explain irregularities or divergences from the fiducial DTD \citep[e.g.][]{Strolger2010,Meng2011,Kistler2013}. Metallicity may affect the observed rate of SNe in two ways: firstly, it can affect the time taken for a star to evolve along the main sequence ($t_{\mathrm{MS}}$) -- Stellar metallicity has varying effects on the main-sequence lifetime of stars \citep[e.g.][]{Georgy2013,Amard2020}, which is also dependent upon initial rotation and degree of mixing; secondly it can affect the time taken from WD formation to SN Ia explosion -- low metallicity stars should produce higher-mass WDs \citep[e.g.][]{Umeda1999,Marigo2007}, resulting in a higher SN Ia rate \citep{Kistler2013}, although note that \citet{Kistler2013} did not find any evidence for this in the data. \citet{Graur2017a} also found that evolution of SN Ia rates with metallicity is consistent with a DTD of the form  $\tau^{-1}$.

The mass-metallicity relation (MZR) is a strong observed correlation between galaxy stellar mass and gas-phase metallicity \citep[e.g.][]{Tremonti2004}, whereby higher mass galaxies have undergone more cycles of stellar evolution and have been polluted with heavy elements created in stars and released in SNe. Thus, the mass of a host galaxy is inextricably linked to the metallicity of its hot gas. However, the nature of the DTD complicates matters, since stars of different ages were formed at different epochs where the galaxy had a different stellar mass and thus different metallicity. Therefore, there is not a direct correlation between the metallicity of SN Ia progenitors and their observed host galaxy stellar mass.  This caveat notwithstanding, it may be an expected consequence of the stipulations of e.g. \citet{Kistler2013} that the rate of SNe Ia is suppressed in the highest mass galaxies where the metallicity is highest, as is the case in our data.

\subsection{Rate per unit stellar mass \label{subsec:snum}}

An alternative view on how the SN Ia rate relates to environment can be gained by calculating the rate of SNe per year per unit stellar mass: SNuM. A measurement of the SNuM is achieved by summing the number of SNe in each stellar mass bin, and dividing by the total stellar mass of field galaxies in that bin. We show the SNuM as a function of galaxy stellar mass in Fig.~ \ref{fig:snum}, along with the measurements from SNe discovered in SDSS spectra \citep{Graur2013} at $z\sim0.1$ and at $z\sim0$ in LOSS as presented in \citet{Li2011a}. We also calculate the SNuM predicuted by our model for each simulated galaxy at a range of redshifts and plot these as curves in Fig.~\ref{fig:snum}. Our model accurately recovers the SNuM measured in DES for the redshift and stellar mass range of the data. While the data from LOSS and SDSS spectra lie slightly higher than our model prediction, the difference is consistent with differences in assumed SFHs, IMFs and stellar population templates as well as SN detection method. Any true residual difference between predicted and observed SNuM at different redshifts could be caused by an evolution of the SN Ia production efficiency as a function of redshift; we defer such an investigation to future work with more refined models and measurements.

There is tentative evidence in the data for a flattening of the SNum vs stellar mass at $\log(M/\mathrm{M}_{\odot})\leq9.5$, which is broadly predicted by the model and is qualitatively consistent with the shape of the analytic fit of \citet{Kistler2013} while standing in slight tension with the conclusions of \citet{Brown2019}. The uncertainty of the measured rates at these low stellar masses prevents us from drawing further conclusions at this stage.
\section{The delay-time distribution as a function of supernova properties}
\label{sec:split_x1_c}

In the previous sections we considered all SNe that passed the various quality, standardisation, and redshift cuts. However, it is well established that SNe Ia show correlations of varying strength between properties of their light curves and host galaxies. For example, it is well established that measures of the light curve duration (e.g. decline rate, or stretch such as SALT2 $x_1$) correlate with host galaxy properties such as morphology \citep{Hamuy1995,Hamuy2000,Mannucci2005}, stellar mass \citep{Kelly2010,Lampeitl2010,Sullivan2010} and specific SFR \citep{Rigault2013,Rigault2018}. If such correlations are caused by intrinsic differences in the progenitors or their surroundings, there may be signatures in the DTDs of SNe when divided into smaller samples. Such a hypothesis was investigated by \citet{Brandt2010} who found evidence for differing DTDs for SNe of low and high stretch, although \citet{Perrett2012} found no evidence for a change in volumetric rate evolution (expected due to the evolution of the CSFR) for the same stretch split. In this section we split the DES SNe Ia by their stretch and colour, and assess how the measured rates and DTDs vary across this parameter space. To do so we split the sample into sub-samples with $x_1 = 0$ and $c=0$ as the divisions points respectively. We repeat the analyses of Sections \ref{sec:rates} and \ref{sec:model} including all of the previous light curve cuts, performing the DTD fitting with a fixed $t_{\mathrm{p}} = 40$~Myr, and present the results in Table \ref{tab:dtd_results_split_lc}. Posterior distributions for fitted parameters are presented in Appendix \ref{appendix:posteriors}.

\subsection{Splitting by SN stretch \label{subsec:split_x1}}

\begin{figure}
    \centering
    \includegraphics[width=.5\textwidth]{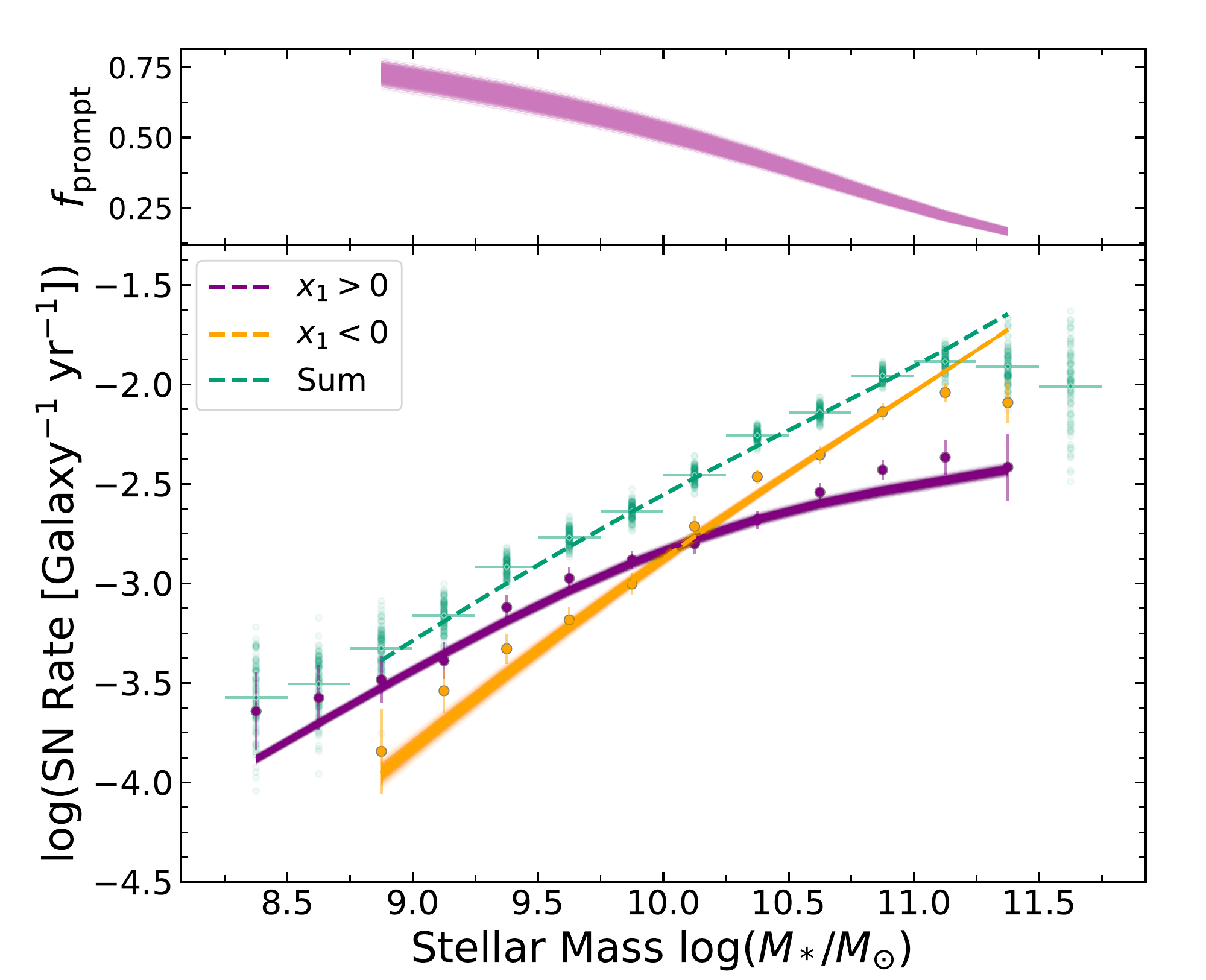}
    \caption{\textit{Upper}: Posterior samples of the fraction of "prompt" SNe resulting from the two-population fit, displayed as a function of host galaxy stellar mass. 
    \textit{Lower}: As per Fig. \ref{fig:rate_fitted_split_x1} but with the data fit simultaneously by a single two-component DTD, with fixed slope $\beta = -1$ and a prompt fraction that is 1 at short delay times, 0 at large delay times, and a linear function of time between $t_1 =0.65$~Gyr (where tardy SNe switch on) and $t_2=0.85$~Gyr (where prompt SNe turn off).
    \label{fig:rate_fitted_split_x1_2c_DTD}}
\end{figure}
The SN Ia rate as a function of stellar mass for SNe split by stretch is shown in Fig. \ref{fig:rate_fitted_split_x1}. The evolution of the rate with stellar mass is significantly different for the two sub-samples: high-stretch SNe dominate the rate at low stellar mass but tail off in the higher mass galaxies, while low-stretch SNe are subdominant in low mass galaxies but display a much steeper dependence on stellar mass and make up the vast majority of SNe in high-mass hosts. These measurements reflect the observed correlations of $x_1$ with stellar mass, sSFR (and by inference, stellar age) as been seen by \citet{Rigault2013,Graur2017,Rigault2018,Rose2019,Hakobyan2020,Nicolas2020,Rose2021}, as well as morphology \citep{Hakobyan2020}. The corresponding DTD fit results in a steep power-law $\beta =-1.70^{+0.19}_{-0.10}$ for high-stretch SNe indicative of a population of predominantly prompt SNe, and a much shallower decay slope ($\beta = -0.79 \pm 0.08$) for low-stretch SNe representing a much more delayed population. The difference in $\beta$ is significant at a level of $3.8\sigma$. This result provides an intriguing confirmation of that observed in 101 SDSS SNe Ia by \citet{Brandt2010} who found that low-$x_1$ SNe displayed much longer delay times than those with high $x_1$.

From the differing DTDs that describe SNe with low and high stretch values we infer that there are either multiple channels through which SN Ia explode which occur on differing timescales (as discussed in \citet{Hakobyan2020}), or that the explosion mechanism evolves with progenitor age. Two scenarios, that are not necessarily mutually exclusive, that could lead to different DTD decay rates are as follows. 
The first such scenario is to assume that all SNe Ia come from the same initial population of stars, and thus that the progenitors of both low and high stretch SNe form at the same time. It is then necessary to invoke models in which the WD progenitors of low-stretch SNe evolve on longer timescales than those of high-stretch SNe, for example due to a different initial binary separation distribution or accretion rate. In this case, the DTD slopes would indeed take on the different values measured here.

Alternatively, low- and high-stretch SNe could form via identical evolutionary channels but with a different onset time since the episode of star-formation. In this scenario, he predominantly high-stretch "prompt" SNe begin exploding as soon as the WDs are formed (nominally 40 Myr) whereas the mainly low-stretch "tardy" SNe only begin exploding after an extended period of time of the order 1 Gyr. The time dependence of both populations would follow the fiducial $\tau^{-1}$ distribution.
Such a scenario may be explained by the sub-$M_{\mathrm{Ch}}$ double-detonation paradigm \citep[e.g.][]{Sim2010,Blondin2017,Shen2017}. Sub-$M_{\mathrm{Ch}}$ SNe Ia typically involve the merger of two WDs and the SN luminosity can be correlated with the mass of the primary WD; a correlation between primary WD mass and age leads to different DTDs of low and high-stretch SNe and hence the different rates observed in this work. Such an evolution of light curve properties with age has recently been seen across the full range of "normal" SNe Ia in 1-dimensional simulations by \citet{Shen2021} and is a promising avenue for further investigation although we note that they struggle to reproduce observed light curves at later times ($t\gtrsim30$~d; \citealt{Shen2021,Gronow2021}).

We explore this scenario by fitting the stretch-separated sub-samples with modified DTDs. We fix the DTD to a power-law with index $-1$, and fix the normalisation to the best fitting value from the full sample. We model the DTD as comprising two components: the prompt, high-stretch SNe and the tardy, low stretch SNe, and we introduce two timescales, $t_1$ and $t_2$. At $\tau\leq t_1$, the DTD is caused solely by prompt SNe; at $\tau \geq t_2$ only tardy SNe explode. In between where $t_1 \leq\tau\leq t_2$, the DTD is a sum of DTD (prompt) and DTD (tardy), where we model the relative fraction $f$ as a smooth linear slope between $t_1$ and $t_2$:
\begin{equation}
    f_{\mathrm{tardy}} = \frac{\tau - t_1}{t_2 - t_1}\,,
\end{equation}
and
\begin{equation}
    f_{\mathrm{prompt}} = 1 - f_{\mathrm{tardy}}\,.
\end{equation}
With $\beta$ and $A$ fixed,  we fit the split-$x_1$ SN Ia rate with $t_1$ and $t_2$ as free parameters, and weak normal priors $p(t_1) \sim \mathcal{N}\left(0.5~\mathrm{Gyr}, 0.5~\mathrm{Gyr}\right)$ $p(t_2) \sim \mathcal{N}\left(1~\mathrm{Gyr}, 0.5~\mathrm{Gyr}\right)$ and the constraint $t_2 > t_1$.

The SN Ia rates generated by the posterior predictive checks are shown in Fig. \ref{fig:rate_fitted_split_x1_2c_DTD}. The best fitting values are mildly consistent with one another, with $t_1 =0.69^{+0.07}_{-0.13}$~Gyr and $t_2 = 0.81^{+0.13}_{-0.10}$~Gyr, corresponding to a relatively sharp transition between the two populations. Our $t_1$ and $t_2$ are reminiscent of the transition found by \citet{Brandt2010} who found high-stretch SNe to be confined to delay times $\lesssim 0.4$~Gyr while low-stretch SNe occurred with delay times $\gtrsim 2.4$~Gyr. The transition found in our analysis appears to occur at much shorter delay times than the several Gyrs found by \citet{Rose2019}. These results are somewhat surprising, as we might expect that the transition is not so sharp but a gradual correlation of delay time with the average stretch of SNe caused by the age vs WD-mass distribution, as predicted by \citet{Shen2021}. However we note that the intrinsic stretch distribution may not be a simple Gaussian, but could instead be bimodal \citep{Scolnic2016,Popovic2021,Nicolas2020}. The correlation of progenitor age and WD mass does not easily reproduce such a bimodal distribution, which may suggest that the sub-$M_{\mathrm{Ch}}$ channel cannot account for all SNe Ia. We also note that our method models the average SFH of galaxies and that the DTD is a statistical distribution: our results do not rule out high-stretch SNe arising from old progenitors, but suggest that such a scenario is unlikely.

\subsubsection{The late end of the DTD \label{subsubsec:subtypes}}

Despite the reasonable fit of the evolving population model, the fit to tardy SNe diverges from the data at high stellar mass, corresponding to the oldest average stellar age. It is difficult to reconcile this turnover with the simple DTD models used thus far in this work, as it would require a steepening or complete turn-off of the DTD at late ($\tau\gtrsim 5$~Gyr) times.

One possibility is that sub-luminous SNe are more numerous in predominantly old stellar populations, such that the fraction passing our light curve cuts is lower than the rest of the sample. Such a phenomenon could be caused by SN Ia sub-classes such as SN 1991bg-like SNe, which are known to explode exclusively in old environments \citep{Perets2010,Barkhudaryan2019} at delay times as long as $>6$~Gyr \citep{Panther2019}, and are typically sub-luminous with low stretch \citep{Gonzalez-Gaitan2014}. If the SN 1991bg-like objects are drawn from the same parent population of WDs as the "normal" SNe Ia, their presence in high-mass hosts could explain the apparent lack of normal SNe Ia in the sample. We check this hypothesis briefly by re-examining the objects left out of our sample due to \texttt{SALT2} cuts. We find that 75\% of objects left out lie in hosts with $M_*>10^{10}$ M$_{\odot}$, while this percentage is 65\% for the objects that pass the cuts. However, of the objects in high-mass galaxies that fail the cuts, only 11 (10\%) fail due to their $x_1$ lying below $-3$, and a total of 20 have $x_1 <-2$ which is the regime populated by SN 1991bg-like objects which is well short of the factor of two by which the DTD model over-predicts the data in the highest mass bin. Another issue is that SN 1991bg-like SNe are generally fainter than normal SNe Ia and the massive galaxies in which they explode are typically bright, leading to a lower detection efficiency than that modelled in Section \ref{subsec:incompleteness_SNe}. A thorough investigation into the presence of SN Ia sub-types in the DES-SN data set is deferred to a future investigation.

\subsection{Splitting by SN colour \label{subsec:split_c}}
\begin{figure}
    \centering
    \includegraphics[width=.5\textwidth]{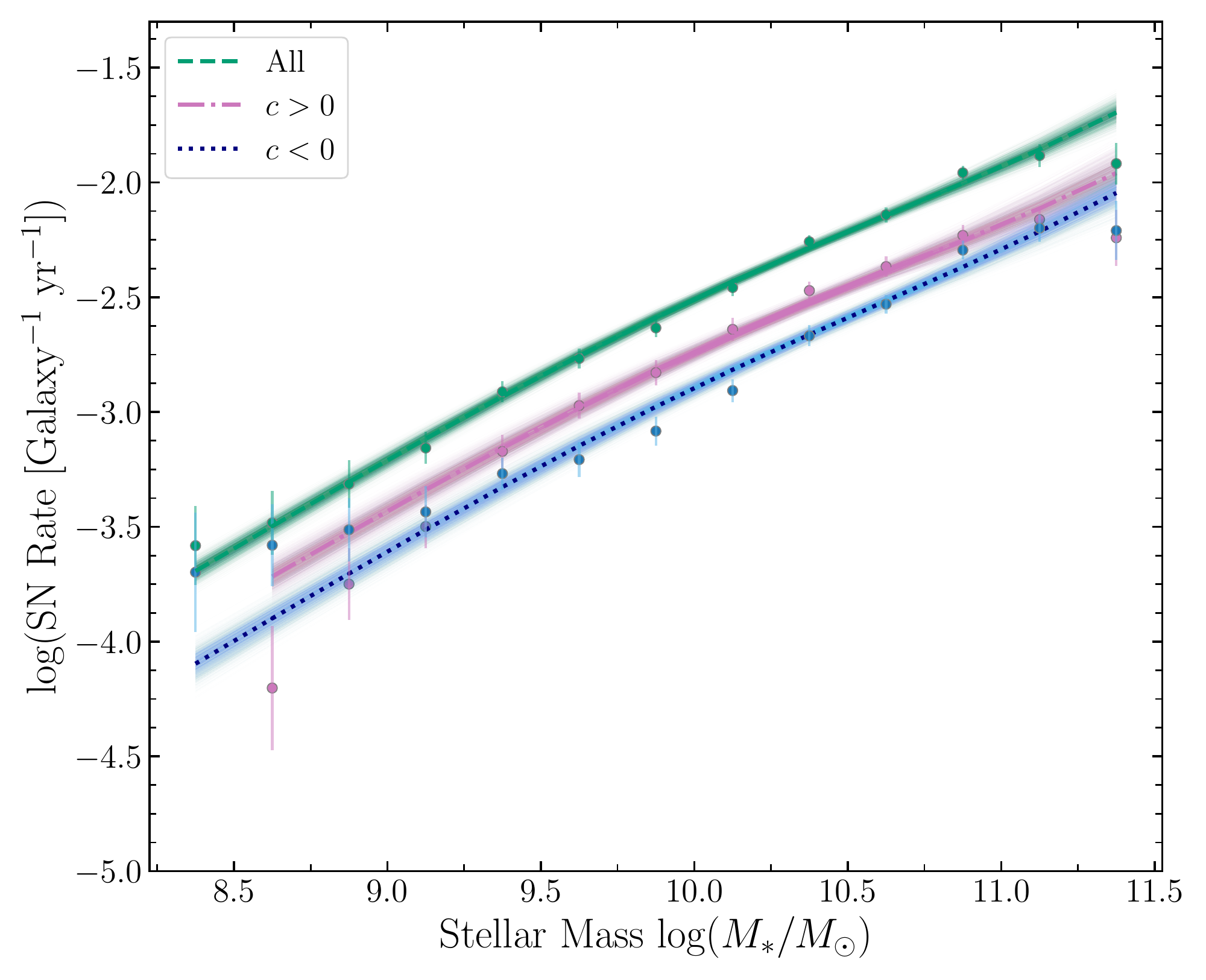}
    \caption{As per Fig. \ref{fig:rate_fitted_split_x1} but with the data split by their colour parameter $c$.
    \label{fig:rate_fitted_split_c}}
\end{figure}

In the previous section, we showed that the slope of the SN Ia rate vs stellar mass relation is different for slow and fast declining supernovae. We find no evidence for a similar difference between SNe with different colours (Fig. \ref{fig:rate_fitted_split_c}), with best fitting values of $\beta$ of $-1.19\pm0.09$ and $-1.09\pm0.08$ for red ($c > 0$) and blue ($c<0$) SNe respectively. This consistency is in agreement with the findings of \citet{Sullivan2010}. Red SNe appear to have a higher normalisation than blue SNe, with roughly 30\% more red than blue SNe. This dominance can be explained by a single population of SNe with an intrinsic colour distribution that is Gaussian and centred on $c=0$ combined with an external dust distribution with an exponential form, systematically shifting the SN colour distribution to the red and causing the higher observed rate of those objects. Such a scenario has been implemented previously in the Multicolor Light-curve Shapes framework \citep{Riess1996,Jha2007}, and recently explored in detail by \citet{Brout2020}. 

In Fig. \ref{fig:rate_fitted_split_c} the predictions from the DTD model differ from the data at low stellar mass ($\log(M_*/\mathrm{M}_{\odot}) <9.25$). Here, the model predicts that red SNe should dominate at the same level as across the full mass range, and the relative difference should remain. The data however become dominated by blue SNe. Although this is where the data is noisiest due to the low number of SNe, it is worth investigating the possible causes of this divergence. A slight preference for SN to be bluer in lower-mass galaxies has been observed \citep{Scolnic2018,Brout2019,Smith2020,Kelsey2021} and could be the same effect seen at low stellar masses here. Physically, the dominance of blue SNe at low stellar masses could either be an intrinsic property of the SNe such as a hotter photosphere in explosions of younger, more massive progenitors, or an extrinsic effect such as a lower average dust column density in low-mass host galaxies. An intermediate explanation is that the SNe in low-mass hosts have a smaller dust column density in the immediate vicinity of the progenitor system, either due to the shorter evolutionary timescale or the higher mass and thus brighter system being less conducive to dust production. A detailed exploration of the effects of environment on SN colour in the DES-SN photometric sample will be presented in Kelsey et al. \textit{in prep}.

\section{Conclusions}
\label{sec:conclusion}

Here, we summarise the key findings presented in this paper:

\begin{itemize}
\item We have measured the rate of SNe Ia per galaxy based on a sample of over 800 SNe and 40,000 galaxies detected by DES. We find a tight linear relationship between SN rate and galaxy stellar mass as seen in SNLS \citep{Sullivan2006} and SDSS \citep{Smith2012} and constrain the slope to high precision. The power-law slope of 0.63 being less than unity indicates that the rate of SNe is not solely driven by stellar mass, but also other factors such as the star-formation history.
    
\item By simulating the stellar mass assembly of average galaxies across cosmic time following the prescription of \citetalias{Childress2014} we have recovered an estimate of the SN Ia DTD, finding a good fit to a power-law distribution with an index of $-1.14 \pm0.05$. Our measurement of the DTD provides a further tight constraint to those of previous measurements \citep[e.g.][]{Graur2013,Maoz2017} and carries signatures predicted by the DD scenario for SNe Ia production.
    
\item We find strong differences in the slope of the SN rate vs stellar mass between SNe with low and high stretch factors ($3.6\sigma$). The differing slopes are readily explained by the correlation between stretch and stellar age, since more massive galaxies play host to older stellar populations which are known to give rise to lower-stretch SNe
    
\item Further investigations of low- and high-stretch SNe reveal two plausible scenarios causing the observed relations. Firstly, low- and high-stretch SNe could belong to separate populations with different DTD slopes, resulting from a different set of initial conditions such as binary separation at the epoch of WD formation. Alternatively, SNe Ia could follow a single DTD where the relative composition of low- and high-stretch objects could change over time. We find a relatively sharp transition at delay times between 0.65 and 0.85~Gyr. Such a scenario is compatible with stretch being related to progenitor mass, a paradigm consistent with a sub-$M_{\mathrm{Ch}}$, double-detonation explosion mechanism. 

\item Red ($c>0$) SNe explode at a higher rate than blue SNe at all mass ranges, we find that DTD decay slope is independent of SN colour. We assume that the higher rate of red SNe is caused by the addition of dust to an intrinsically Gaussian colour distribution centred slightly bluewards of $c=0$, but defer a detailed investigation to future work.
\end{itemize}

\section*{Software}

All software used in this publication is publicly available. In particular, we made extensive use of \texttt{numpy} \citep{Harris2020}, \texttt{Astropy} \citep{AstropyCollaboration2013,AstropyCollaboration2018}, \texttt{matplotlib} \citep{Hunter2007}, \texttt{SciPy} \citep{Virtanen2020}, \texttt{pandas} \citep{Mckinney2010}, \texttt{Stan} \citep{Carpenter2017}, \texttt{seaborn} \citep{Waskom2020}, and \texttt{ChainConsumer} \citep{Hinton2016}.

\section*{Acknowledgements}
We thank J. Michael Burgess for discussions regarding model fitting in Stan.
We thank the reviewer, Mickael Rigault, for constructive comments on the manuscript.

 P.W. acknowledges support from the Science and Technology Facilities Council (STFC) grant ST/R000506/1. M.Su. and M.Sm. acknowledge support from EU/FP7-ERC grant 615929. MS is funded by the European Reearch Council (ERC) under the European Union's Horizon 2020 Research and Innovation program (grant agreement no 759194 - USNAC). L.G. acknowledges financial support from the Spanish Ministry of Science, Innovation and Universities (MICIU) under the 2019 Ram\'on y Cajal program RYC2019-027683 and from the Spanish MICIU project PID2020-115253GA-I00.
  
Funding for the DES Projects has been provided by the U.S. Department of Energy, the U.S. National Science Foundation, the Ministry of Science and Education of Spain, 
the Science and Technology Facilities Council of the United Kingdom, the Higher Education Funding Council for England, the National Center for Supercomputing 
Applications at the University of Illinois at Urbana-Champaign, the Kavli Institute of Cosmological Physics at the University of Chicago, 
the Center for Cosmology and Astro-Particle Physics at the Ohio State University,
the Mitchell Institute for Fundamental Physics and Astronomy at Texas A\&M University, Financiadora de Estudos e Projetos, 
Funda{\c c}{\~a}o Carlos Chagas Filho de Amparo {\`a} Pesquisa do Estado do Rio de Janeiro, Conselho Nacional de Desenvolvimento Cient{\'i}fico e Tecnol{\'o}gico and 
the Minist{\'e}rio da Ci{\^e}ncia, Tecnologia e Inova{\c c}{\~a}o, the Deutsche Forschungsgemeinschaft and the Collaborating Institutions in the Dark Energy Survey. 

The Collaborating Institutions are Argonne National Laboratory, the University of California at Santa Cruz, the University of Cambridge, Centro de Investigaciones Energ{\'e}ticas, 
Medioambientales y Tecnol{\'o}gicas-Madrid, the University of Chicago, University College London, the DES-Brazil Consortium, the University of Edinburgh, 
the Eidgen{\"o}ssische Technische Hochschule (ETH) Z{\"u}rich, 
Fermi National Accelerator Laboratory, the University of Illinois at Urbana-Champaign, the Institut de Ci{\`e}ncies de l'Espai (IEEC/CSIC), 
the Institut de F{\'i}sica d'Altes Energies, Lawrence Berkeley National Laboratory, the Ludwig-Maximilians Universit{\"a}t M{\"u}nchen and the associated Excellence Cluster Universe, 
the University of Michigan, NFS's NOIRLab, the University of Nottingham, The Ohio State University, the University of Pennsylvania, the University of Portsmouth, 
SLAC National Accelerator Laboratory, Stanford University, the University of Sussex, Texas A\&M University, and the OzDES Membership Consortium.

Based in part on observations at Cerro Tololo Inter-American Observatory at NSF's NOIRLab (NOIRLab Prop. ID 2012B-0001; PI: J. Frieman), which is managed by the Association of Universities for Research in Astronomy (AURA) under a cooperative agreement with the National Science Foundation.

The DES data management system is supported by the National Science Foundation under Grant Numbers AST-1138766 and AST-1536171.
The DES participants from Spanish institutions are partially supported by MICINN under grants ESP2017-89838, PGC2018-094773, PGC2018-102021, SEV-2016-0588, SEV-2016-0597, and MDM-2015-0509, some of which include ERDF funds from the European Union. IFAE is partially funded by the CERCA program of the Generalitat de Catalunya.
Research leading to these results has received funding from the European Research
Council under the European Union's Seventh Framework Program (FP7/2007-2013) including ERC grant agreements 240672, 291329, and 306478.
We  acknowledge support from the Brazilian Instituto Nacional de Ci\^encia
e Tecnologia (INCT) do e-Universo (CNPq grant 465376/2014-2).

This manuscript has been authored by Fermi Research Alliance, LLC under Contract No. DE-AC02-07CH11359 with the U.S. Department of Energy, Office of Science, Office of High Energy Physics.

\section*{Data Availability}

The DES-SN photometric SN Ia catalogue and associated host galaxy data will be made available as part of the DES5YR SN cosmology analysis at https://des.ncsa.illinois.edu/releases/sn. Field galaxy data including photometric redshifts will appear alongside other data products released as part of the DES Y3 weak lensing cosmology analysis at https://des.ncsa.illinois.edu/releases.



\bibliographystyle{mnras}
\bibliography{PhilMendeley} 




\appendix

\section{Linear fits using Bayesian inference}
In this section we describe the procedures used to fit slopes and intercepts to the relationships measured in the analysis.

\label{appendix:linear_fits}

\begin{figure}
    \centering
    \includegraphics[width=0.5\textwidth]{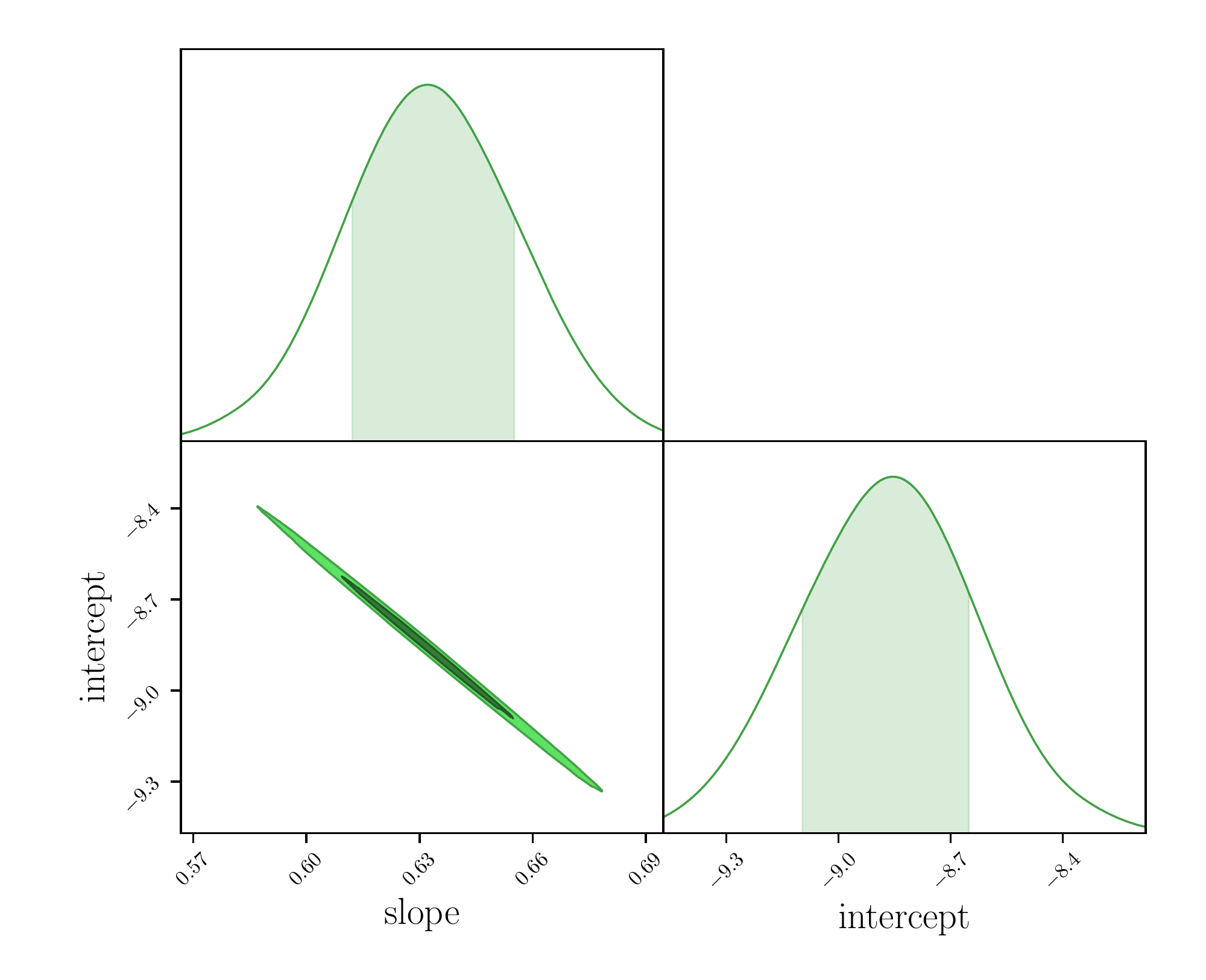}
    \caption{Joint posterior distribution for the slope and intercept of the linear fit to the SN rate per galaxy per year as a function of stellar mass (Section \ref{sec:rates}; Fig. \ref{fig:rate_fitted}).}
    \label{fig:corner_slope_int}
\end{figure}

We model the relation with the linear relationship: 
\begin{equation}
    R_{\mathrm{G}} = \frac{\mathrm{d}R}{\mathrm{d}M_*} M_* + c \,,
\label{eq:rate_fit}
\end{equation}
where $dR/dM_*$ signifies the change of the rate of SNe as a function of the stellar mass, $c$ is a constant that sets the normalisation of the rate. We fit the model assuming a normal likelihood, such that the observed rate in each stellar mass bin is itself modelled as a Gaussian distribution described by the mean and standard deviation of the data in that bin. We adopt weakly informative normal priors on the slope and intercept: $p\left(\frac{\mathrm{d}R}{\mathrm{d}M_*}\right) \sim \mathcal{N}\left(0,5\right)$ and  $p\left(c\right) \sim \mathcal{N}\left(-12,5\right)$ respectively. We sample using 4 chains, each with 2000 warm up and 2000 sampling iterations. We report parameter estimates based on the mean and standard deviation of their posterior samples. We use the $\hat{R}$ diagnostic of \citep{Vehtari2019} to assess the convergence of MC chains, and only accept fits where $\hat{R}<1.05$.

The joint posterior distribution for the slope and intercept of the overall SN rate vs stellar mass are shown in Fig. \ref{fig:corner_slope_int}. The two parameters are highly degenerate, yet well constrained. This degeneracy manifests as the spread of potential linear fits as drawn in light green on Fig. \ref{fig:rate_fitted}.

\section{Posterior distributions for DTD fits}
\label{appendix:posteriors}

\begin{figure}
    \centering
    \includegraphics[width=.5\textwidth]{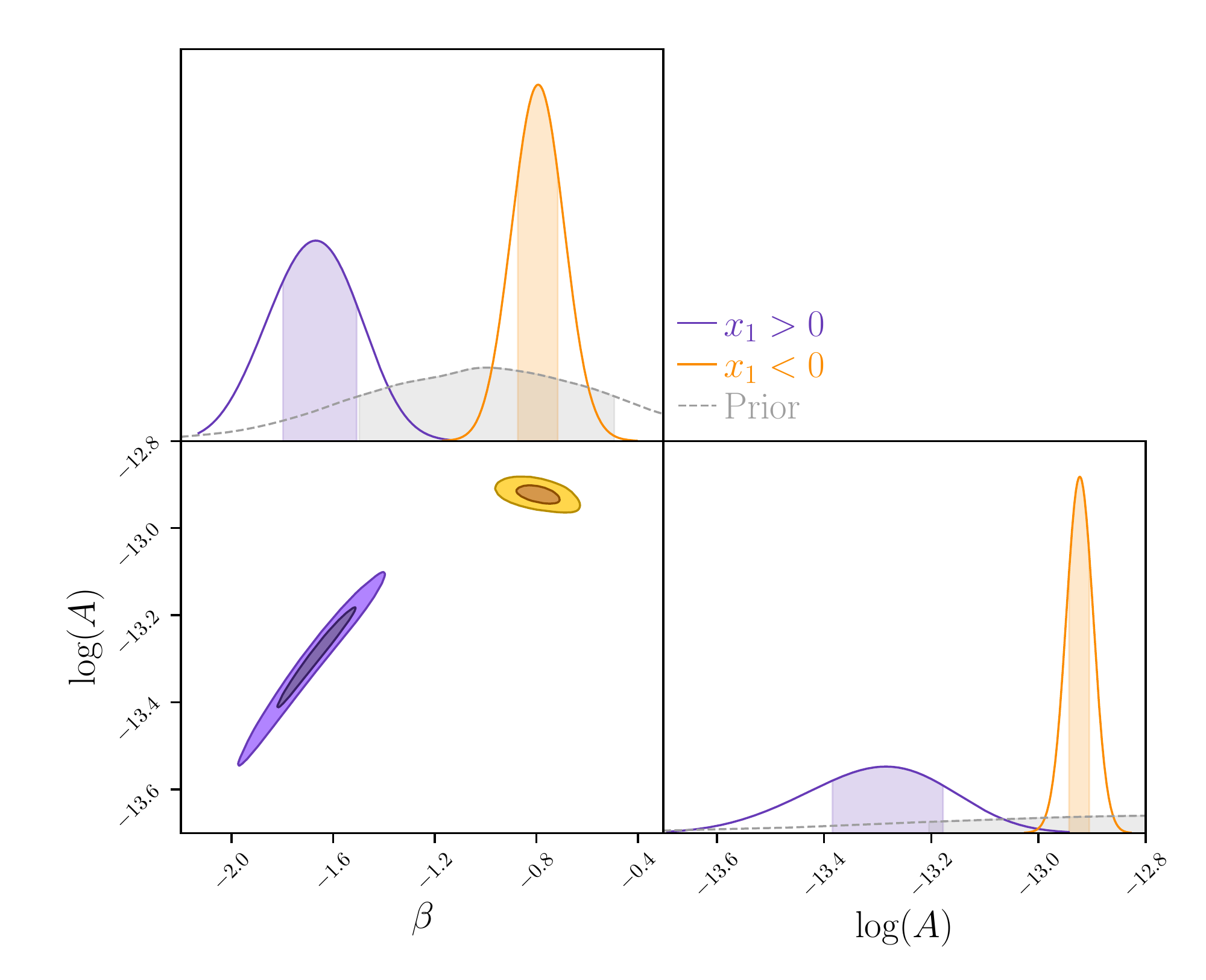}
    \caption{Posterior distributions for DTD slope $\beta$ and normalisation $A$ for SNe Ia split by their $x_1$ parameter.
    \label{fig:corner_beta_norm_split_x1}}
\end{figure}

\begin{figure}
    \centering
    \includegraphics[width=.5\textwidth]{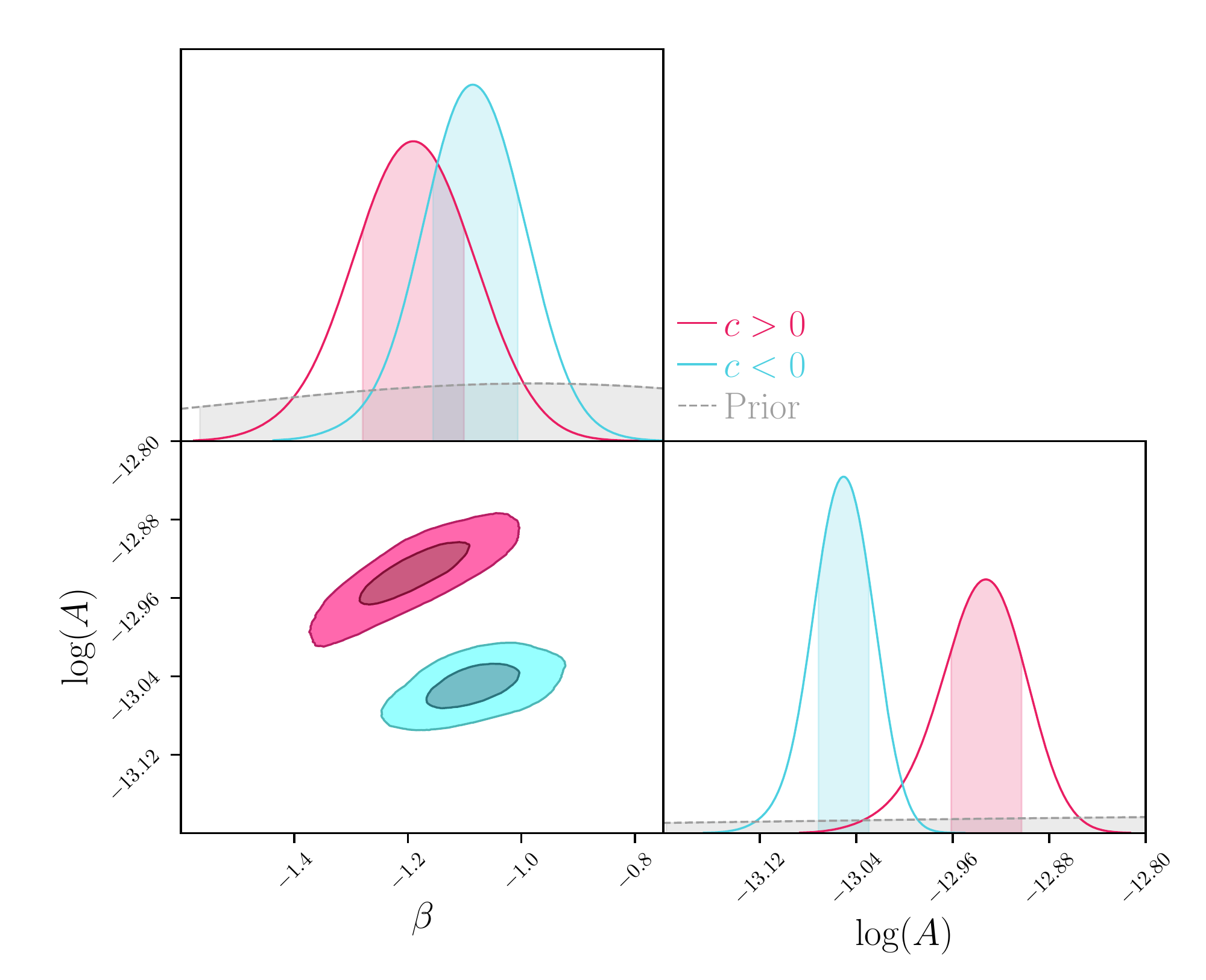}
    \caption{Posterior distributions for DTD slope $\beta$ and normalisation $A$ for SNe Ia split by their $c$ parameter.
    \label{fig:corner_beta_norm_split_c}}
\end{figure}
\begin{figure}
    \centering
    \includegraphics[width=.5\textwidth]{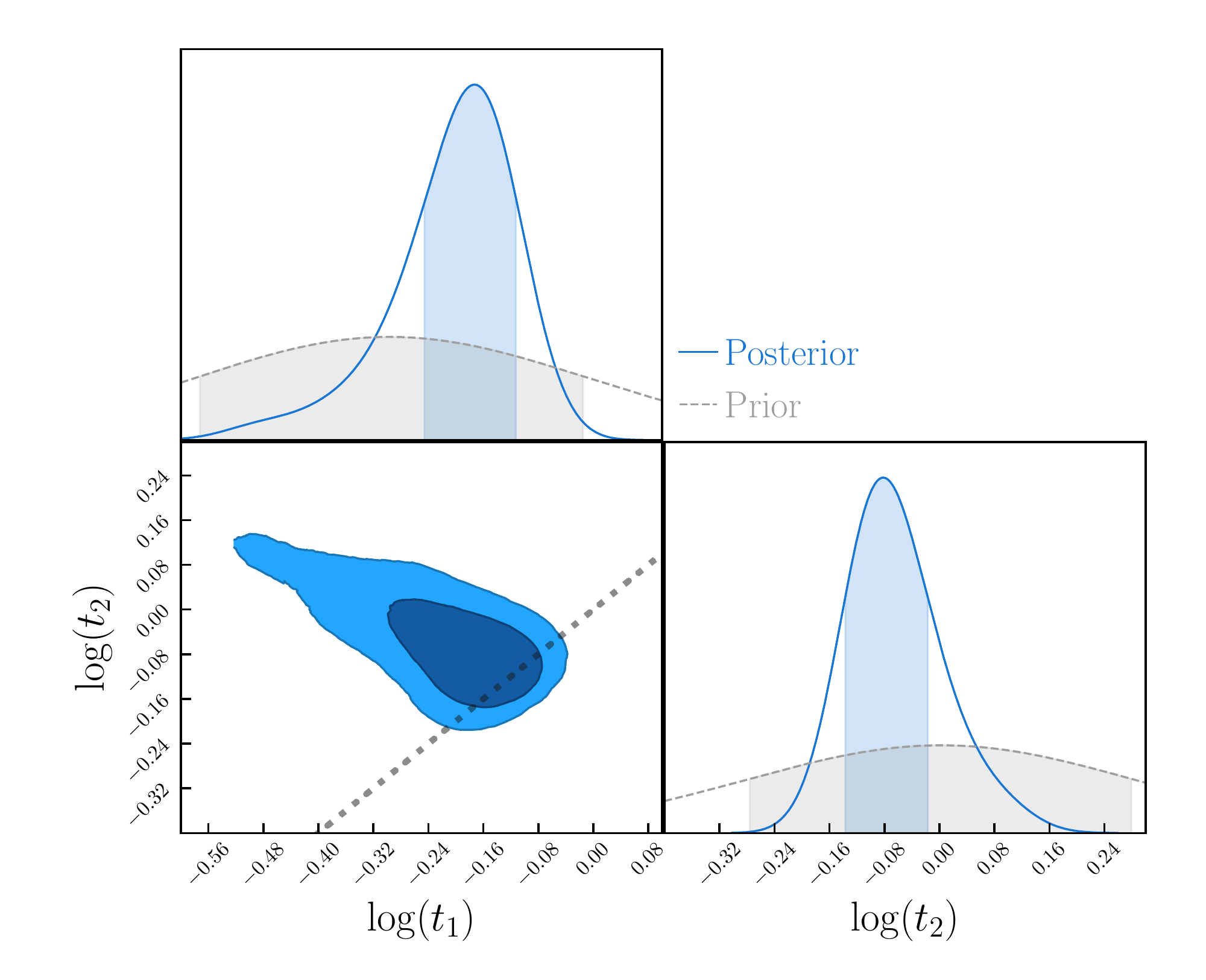}
    \caption{Posterior distributions for $t_1$ and $t_2$ of the model presented in Section \ref{subsec:split_x1}.
    \label{fig:corner_t1_t2}}
\end{figure}
In this section we present the joint and marginal posterior distributions for the various fits of the DTD presented in Section \ref{sec:split_x1_c}. Posteriors have been displayed using v0.33.0 of \texttt{ChainConsumer} \citep{Hinton2016}.
\clearpage
\onecolumn
\parbox{\textwidth}{
\scriptsize
$^{1}$ School of Physics and Astronomy, University of Southampton,  Southampton, SO17 1BJ, UK\\
$^{2}$ Univ Lyon, Univ Claude Bernard Lyon 1, CNRS, IP2I Lyon / IN2P3, IMR 5822, F-69622, Villeurbanne, France\\
$^{3}$ Institute of Cosmology and Gravitation, University of Portsmouth, Portsmouth, PO1 3FX, UK\\
$^{4}$ Department of Physics, Duke University Durham, NC 27708, USA\\
$^{5}$ The Research School of Astronomy and Astrophysics, Australian National University, ACT 2601, Australia\\
$^{6}$ NASA Einstein Fellow\\
$^{7}$ Center for Astrophysics $\vert$ Harvard \& Smithsonian, 60 Garden Street, Cambridge, MA 02138, USA\\
$^{8}$ School of Mathematics and Physics, University of Queensland,  Brisbane, QLD 4072, Australia\\
$^{9}$ Institute of Space Sciences (ICE, CSIC), Campus UAB, Carrer de Can Magrans, s/n, E-08193 Barcelona, Spain.\\
$^{10}$ Department of Astronomy and Astrophysics, University of Chicago, Chicago, IL 60637, USA\\
$^{11}$ Kavli Institute for Cosmological Physics, University of Chicago, Chicago, IL 60637, USA\\
$^{12}$ Centre for Gravitational Astrophysics, College of Science, The Australian National University, ACT 2601, Australia\\
$^{13}$ Universit\'e Clermont Auvergne, CNRS/IN2P3, LPC, F-63000 Clermont-Ferrand, France\\
$^{14}$ Centro de Investigaciones Energ\'eticas, Medioambientales y Tecnol\'ogicas (CIEMAT), Madrid, Spain\\
$^{15}$ INAF, Osservatorio Astronomico di Trieste, I-34143 Trieste, Italy\\
$^{16}$ Centre for Astrophysics \& Supercomputing, Swinburne University of Technology, Victoria 3122, Australia\\
$^{17}$ Sydney Institute for Astronomy, School of Physics, A28, The University of Sydney, NSW 2006, Australia\\
$^{18}$ Cerro Tololo Inter-American Observatory, NSF's National Optical-Infrared Astronomy Research Laboratory, Casilla 603, La Serena, Chile\\
$^{19}$ Laborat\'orio Interinstitucional de e-Astronomia - LIneA, Rua Gal. Jos\'e Cristino 77, Rio de Janeiro, RJ - 20921-400, Brazil\\
$^{20}$ Fermi National Accelerator Laboratory, P. O. Box 500, Batavia, IL 60510, USA\\
$^{21}$ Instituto de F\'{i}sica Te\'orica, Universidade Estadual Paulista, S\~ao Paulo, Brazil\\
$^{22}$ CNRS, UMR 7095, Institut d'Astrophysique de Paris, F-75014, Paris, France\\
$^{23}$ Sorbonne Universit\'es, UPMC Univ Paris 06, UMR 7095, Institut d'Astrophysique de Paris, F-75014, Paris, France\\
$^{24}$ Department of Physics \& Astronomy, University College London, Gower Street, London, WC1E 6BT, UK\\
$^{25}$ Kavli Institute for Particle Astrophysics \& Cosmology, P. O. Box 2450, Stanford University, Stanford, CA 94305, USA\\
$^{26}$ SLAC National Accelerator Laboratory, Menlo Park, CA 94025, USA\\
$^{27}$ Instituto de Astrofisica de Canarias, E-38205 La Laguna, Tenerife, Spain\\
$^{28}$ Universidad de La Laguna, Dpto. Astrofísica, E-38206 La Laguna, Tenerife, Spain\\
$^{29}$ Center for Astrophysical Surveys, National Center for Supercomputing Applications, 1205 West Clark St., Urbana, IL 61801, USA\\
$^{30}$ Department of Astronomy, University of Illinois at Urbana-Champaign, 1002 W. Green Street, Urbana, IL 61801, USA\\
$^{31}$ Institut de F\'{\i}sica d'Altes Energies (IFAE), The Barcelona Institute of Science and Technology, Campus UAB, 08193 Bellaterra (Barcelona) Spain\\
$^{32}$ Astronomy Unit, Department of Physics, University of Trieste, via Tiepolo 11, I-34131 Trieste, Italy\\
$^{33}$ INAF-Osservatorio Astronomico di Trieste, via G. B. Tiepolo 11, I-34143 Trieste, Italy\\
$^{34}$ Institute for Fundamental Physics of the Universe, Via Beirut 2, 34014 Trieste, Italy\\
$^{35}$ Observat\'orio Nacional, Rua Gal. Jos\'e Cristino 77, Rio de Janeiro, RJ - 20921-400, Brazil\\
$^{36}$ Department of Physics, University of Michigan, Ann Arbor, MI 48109, USA\\
$^{37}$ Department of Physics, IIT Hyderabad, Kandi, Telangana 502285, India\\
$^{38}$ Santa Cruz Institute for Particle Physics, Santa Cruz, CA 95064, USA\\
$^{39}$ Institute of Theoretical Astrophysics, University of Oslo. P.O. Box 1029 Blindern, NO-0315 Oslo, Norway\\
$^{40}$ Institut d'Estudis Espacials de Catalunya (IEEC), 08034 Barcelona, Spain\\
$^{41}$ Institute of Space Sciences (ICE, CSIC),  Campus UAB, Carrer de Can Magrans, s/n,  08193 Barcelona, Spain\\
$^{42}$ Instituto de Fisica Teorica UAM/CSIC, Universidad Autonoma de Madrid, 28049 Madrid, Spain\\
$^{43}$ Institute of Astronomy, University of Cambridge, Madingley Road, Cambridge CB3 0HA, UK\\
$^{44}$ Kavli Institute for Cosmology, University of Cambridge, Madingley Road, Cambridge CB3 0HA, UK\\
$^{45}$ Department of Physics, Stanford University, 382 Via Pueblo Mall, Stanford, CA 94305, USA\\
$^{46}$ Center for Cosmology and Astro-Particle Physics, The Ohio State University, Columbus, OH 43210, USA\\
$^{47}$ Department of Physics, The Ohio State University, Columbus, OH 43210, USA\\
$^{48}$ Faculty of Physics, Ludwig-Maximilians-Universit\"at, Scheinerstr. 1, 81679 Munich, Germany\\
$^{49}$ Max Planck Institute for Extraterrestrial Physics, Giessenbachstrasse, 85748 Garching, Germany\\
$^{50}$ Department of Astronomy/Steward Observatory, University of Arizona, 933 North Cherry Avenue, Tucson, AZ 85721-0065, USA\\
$^{51}$ Australian Astronomical Optics, Macquarie University, North Ryde, NSW 2113, Australia\\
$^{52}$ Lowell Observatory, 1400 Mars Hill Rd, Flagstaff, AZ 86001, USA\\
$^{53}$ George P. and Cynthia Woods Mitchell Institute for Fundamental Physics and Astronomy, and Department of Physics and Astronomy, Texas A\&M University, College Station, TX 77843,  USA\\
$^{54}$ Department of Astronomy, The Ohio State University, Columbus, OH 43210, USA\\
$^{55}$ Radcliffe Institute for Advanced Study, Harvard University, Cambridge, MA 02138\\
$^{56}$ Instituci\'o Catalana de Recerca i Estudis Avan\c{c}ats, E-08010 Barcelona, Spain\\
$^{57}$ Physics Department, 2320 Chamberlin Hall, University of Wisconsin-Madison, 1150 University Avenue Madison, WI  53706-1390\\
$^{58}$ Department of Astrophysical Sciences, Princeton University, Peyton Hall, Princeton, NJ 08544, USA\\
$^{59}$ Department of Physics and Astronomy, Pevensey Building, University of Sussex, Brighton, BN1 9QH, UK\\
$^{60}$ Computer Science and Mathematics Division, Oak Ridge National Laboratory, Oak Ridge, TN 37831\\
$^{61}$ Universit\"ats-Sternwarte, Fakult\"at f\"ur Physik, Ludwig-Maximilians Universit\"at M\"unchen, Scheinerstr. 1, 81679 M\"unchen, Germany\\
}

\bsp	
\label{lastpage}
\end{document}